\documentclass[aps,pre,twocolumn,superscriptaddress,showpacs]{revtex4-2}
\usepackage{graphicx}
\usepackage{array}
\usepackage{amsfonts}
\usepackage{amssymb,amsmath,multirow,rotate}
\usepackage{mathrsfs}
\usepackage{booktabs}
\usepackage{threeparttable}
\usepackage{multirow}
\usepackage{epsfig}
\usepackage{threeparttable}
\usepackage{chngpage}
\usepackage{latexsym}
\usepackage{dcolumn}
\usepackage{graphics,graphicx,bm,fleqn,epic,eepic,float}
\usepackage{verbatim}
\usepackage[dvipsnames]{xcolor}
\usepackage{xr}
\usepackage{float}
\usepackage{listings} 
\usepackage{placeins} 
\usepackage{tikz}
\usepackage{amsmath, nccmath}
\usepackage{soul}

\definecolor{red}{rgb}{1,0,0}

\definecolor{blue}{rgb}{0,0,1}

\begin{document}

\title{Unsupervised learning for anticipating critical transitions}

\date{\today}

\author{Shirin Panahi}
\affiliation{School of Electrical, Computer, and Energy Engineering, Arizona State University, Tempe, AZ 85287, USA}

\author{Ling-Wei Kong}
\affiliation{Department of Computational Biology, Cornell University, Ithaca, NY 14850, USA}

\author{Bryan Glaz}
\affiliation{Vehicle Technology Directorate, CCDC Army Research Laboratory, 2800 Powder Mill Road, Adelphi, MD 20783-1138, USA}

\author{Mulugeta Haile}
\affiliation{Vehicle Technology Directorate, CCDC Army Research Laboratory, 6340 Rodman Road, Aberdeen Proving Ground, MD 21005-5069, USA}

\author{Ying-Cheng Lai} \email{Ying-Cheng.Lai@asu.edu}
\affiliation{School of Electrical, Computer, and Energy Engineering, Arizona State University, Tempe, AZ 85287, USA}
\affiliation{Department of Physics, Arizona State University, Tempe, Arizona 85287, USA}

\begin{abstract}


For anticipating critical transitions in complex dynamical systems, the recent approach of parameter-driven reservoir computing requires explicit knowledge of the bifurcation parameter. We articulate a framework combining a variational autoencoder (VAE) and reservoir computing to address this challenge. In particular, the driving factor is detected from time series using the VAE in an unsupervised-learning fashion and the extracted information is then used as the parameter input to the reservoir computer for anticipating the critical transition. We demonstrate the power of the unsupervised learning scheme using prototypical dynamical systems including the spatiotemporal Kuramoto–Sivashinsky system. The scheme can also be extended to scenarios where the target system is driven by several independent parameters or with partial state observations.

\end{abstract}

\maketitle

In complex dynamical systems ranging from ecological~\cite{hastings2018transient} 
and climate systems~\cite{lenton2019climate} to infrastructure~\cite{DC:1989} and 
social~\cite{Scheffer2020,o2020transient} networks, critical transitions can occur 
in which the system undergoes an abrupt and often catastrophic switch to a 
characteristically different final state~\cite{NASBFBS:2016,Scheffer2020}. 
Anticipating critical transitions is important to disaster prediction and 
prevention, risk mitigation, and resilience 
enhancement~\cite{Schefferetal:2009,Scheffer2012}. If an accurate model of the 
system is available with the knowledge of the bifurcation parameter and its 
possible variation into the future, computations can be carried out to anticipate 
a critical transition. The typical real-world scenario is that the mathematical 
model of the underlying system is not available, so one must rely on observational 
or measurement data to anticipate critical transitions but this can be quite 
challenging. 

There are two main approaches to data-based anticipation of critical 
transitions [Sec.~S1 in Supplementary Information (SI)~\cite{SI}]. One is 
based on finding the system equations from data using sparse 
optimization~\cite{WYLKG:2011}, which is effective if the equation structure of 
the underlying system is particularly simple. The second approach is based on machine 
learning~\cite{Lim:2020,BSPSLAB:2021,KFGL:2021a,KLNPB:2021,FKLW:2021,KFGL:2021b,KWGHL:2023}. 
A recent method~\cite{KFGL:2021a,KWGHL:2023} is parameter-driven reservoir 
computing~\cite{Jaeger:2001,MNM:2002} for nonlinear and complex dynamical
systems~\cite{PHGLO:2018,Bollt:2021,GBGB:2021,ZKL:2023,KBL:2024,YHBTLS:2024}.
Specifically, in conventional reservoir computing, 
the neural network is trained to learn the ``dynamical climate'' of the target 
system. However, parameter-driven reservoir computing also enables the neural 
network to learn how the dynamical climate changes with a system 
parameter~\cite{KFGL:2021a,KWGHL:2023}, which was applied to anticipating 
crises~\cite{KFGL:2021a}, abrupt onset or destruction of synchronization in 
coupled oscillator systems~\cite{FKLW:2021}, amplitude death~\cite{XKSL:2021}, 
and the occurrence of periodical windows~\cite{PCGPO:2021}. Parameter-driven 
reservoir computing has also been exploited for constructing digital twins of 
nonlinear systems~\cite{KWGHL:2023} and for anticipating tipping in real-world 
systems such as the potential collapse of Atlantic Meridional Overturning 
Circulation~\cite{Panahi2024}. A limitation is that the knowledge of the 
time-dependent bifurcation parameter is required.

In this Letter, we develop a machine-learning framework that combines a 
variational autoencoder (VAE) with parameter-driven reservoir computing to 
address the challenge of anticipating critical transitions without explicit 
knowledge of the bifurcation parameter. In particular, we use VAE to extract the 
parameter that drives the system towards a critical point from the available time 
series data. This is essentially {\em unsupervised} learning leading to a 
``parameter'' of the system and its variations. With this information, 
parameter-driven reservoir computing is then exploited to anticipate a critical 
transition. We demonstrate the power of the unsupervised learning scheme using 
prototypical nonlinear dynamical systems. We also generalize the framework to 
scenarios where the target system is driven by more than one independent 
parameter or only partial state observation is available. While the previous 
parameter-driven reservoir-computing scheme is data-driven and requires no system 
models, it still needs knowledge about the bifurcation parameter whose variations 
leads to a critical transition. Our unsupervised-learning framework relaxes this
requirement in that no prior knowledge of the system parameter is needed. In fact, 
the bifurcation parameter and its variations can be faithfully extracted from 
data. This makes machine-learning based prediction of critical transitions a 
step closer to real applications.

\begin{figure} [ht!]
\centering
\includegraphics[width=0.8\linewidth]{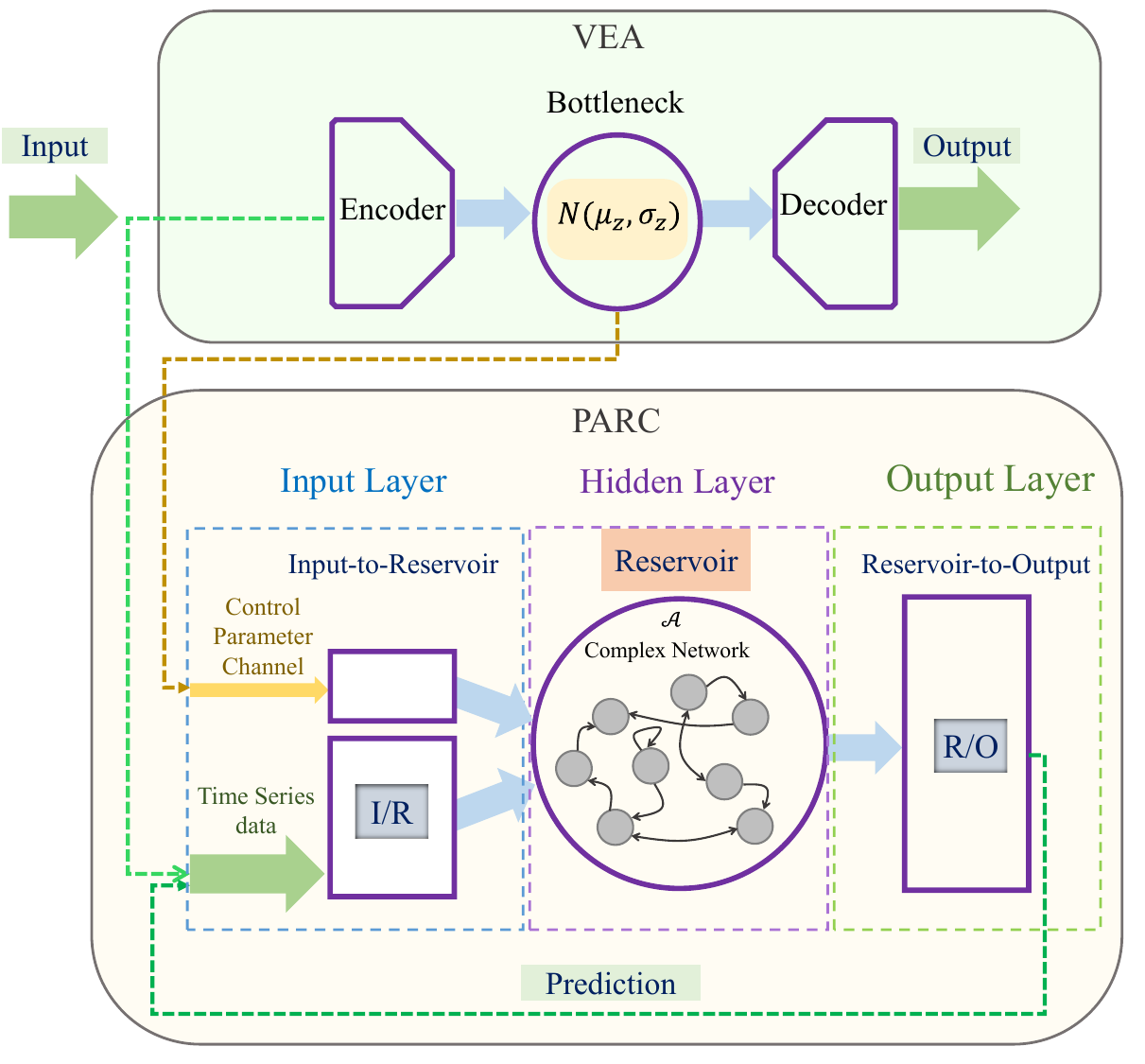}
\caption{Integrated architecture of VAE and parameter-driven reservoir computing. 
The VAE consists of the encoder and the decoder, with the latent distribution 
parameter extracted from the input data in the bottleneck, and it identifies the 
bifurcation parameter and extracts its variations. The input to the reservoir 
computer consist of the time-series data from the target system and the VAE-extracted 
parameter variations, and the output is the prediction of long-term dynamics.}
\label{fig:Archi}
\end{figure}

Figure~\ref{fig:Archi} outlines the main features of our proposed architecture of
unsupervised leaning for anticipating critical transitions, tailored to scenarios 
where the bifurcation parameter is not accessible. A VAE is first trained to 
extract the parameters characterizing the system dynamics and provide multi-step 
prediction. In particular, the VAE has an encoder and a decoder. When presented 
with a time-series dataset, the encoder extracts the latent parameters generating 
the dynamic variations in the dataset, and the decoder employs the latent 
parameters encapsulating essential information about the dynamics to reconstruct 
the time series from the given initial condition. The trained VAE serves as the 
input parameter component in parameter-driven reservoir computing.  

Since the purpose of the VAE is processing time series to extract information about
the parameter variations, we use a deep convolution neural network (DCNN) as the 
encoder. The DCNN detects dynamical features from the input time series and 
outputs $\mu$ and $\sigma$ that parameterize the Gaussian distribution of the 
bifurcation parameter $z$ for each input $x$. For the decoder, we employ a 
feedforward neural network (FNN) that propagates a given initial condition $x_0$ 
for several time steps through a feedback loop. Multiple-step predictions 
$\hat{x}$ are generated via iterative single-step prediction by the decoder 
FNN, where the output of the last one-step prediction $\hat{x}$ becomes the 
input for the next iteration. All iterations are also modulated by the extracted 
parameter $z$. That is, the decoder works in a similar fashion to parameter-driven 
reservoir computing, but with a simpler structure that allows back-propagation 
through the entire VAE. The initial value $x_0$ is chosen from a random training 
sample $x$. The training goal of the VAE is reconstructing the time series 
immediately after $x_0$ for a certain length. Overall, during the training phase, 
the VAE is optimized to align the output of the decoder with a time-series example 
from the normal region of the dataset. The overarching objective of the VAE 
architecture is enabling the decoder to influence the encoder in extracting a 
small number of meaningful and informative latent parameters. After the training, 
this collaborative process ensures that the VAE identifies the bifurcation 
parameters from the time-series data at the VAE's information bottleneck 
[Sec.~S2 in SI~\cite{SI}).

More specifically, the encoder uses the input series $x$ to provide a normal 
distribution as an output for each latent parameter $z$ with mean $\mu_{z}$ and 
variance $\sigma^2_{z}$. For the training, samples of each latent parameter $z$ 
are taken according to $z = \mu_z + \sigma_z \epsilon$ from the distribution 
$\mathcal{N}(\mu_z,\sigma^2_z)$ where $\epsilon \sim \mathcal{N}(0, 1)$ is 
independently sampled for every training example during each training step. The 
decoder uses these samples along with an initial condition $x_0$ of the target 
system for predicting its time series up to an arbitrary future time $T$ into the
future. Providing the decoder with a randomly sampled initial condition ensures 
the focus of the encoder on deciphering the parameters characterizing the dynamics 
of the data rather than encoding a particular state of the system. The end-to-end 
training process contains: (1) a mean-squared error loss between the predicted 
propagation series $\hat{x}_p$ from the decoder and the target time series $x$, 
and (2) the VAE regularization term $R_{z} = E_1 + E_2 + E_3$, with the following
meanings of the three terms. The first term $E_1 = D_{KL}[q(z,x)||q(z)p_D(x)]$ is 
the mutual information among the latent variables and the input data, where 
$D_{KL}$ denotes the Kullback-Leibler (KL) divergence, $p_D (x)$ is the data 
distribution, and $q(z,x)$ is the joint distribution of the latent parameters 
and the data. The second term $E_2 = D_{KL}[q(z)||\prod_i q(z_i)]$ measures the 
total correlation among the latent parameters [$q(z) = \int dx q(z,x)$, with 
$q(z_i)=\int dx \Pi_{i\neq j} dz_j q(z,x)$]. The third term 
$E_3 = \sum_i D_{KL}[q(z_i)||p(z_i)]$ regularizes the latent space 
[$p(z_i)=\mathcal{N}(0, 1)$], ensuring that each latent variable follows a 
standard normal distribution to avoid overfitting and maintain generalizability. 
After the VAE is trained, the input time series $x$ and the detected latent 
variable $z$ as the control parameters are used as input for training the 
parameter-driven reservoir computer. For evaluation, the output prediction of 
the reservoir computer is monitored for any change in the dynamics and possible 
system collapse by changing the latent parameter beyond the VAE extracted regime
[see Sec.~S3 in SI~\cite{SI} for details of parameter-driven reservoir computing].

\begin{figure} [ht!]
\centering
\includegraphics[width=\linewidth]{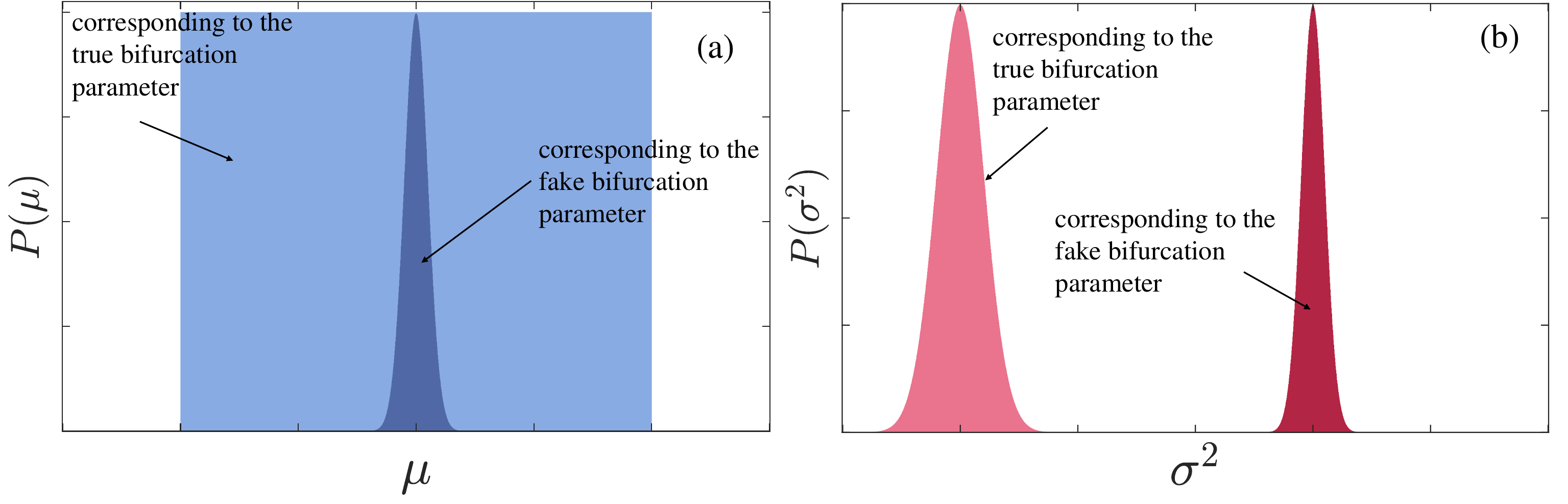}
\caption{Principle of VAE identification of the bifurcation parameters. The VAE 
designates latent-parameter channels whose number is larger than the actual 
number of the bifurcation parameters. With the time series data from distinct 
bifurcation-parameter values as the input to the VAE, each latent parameter 
channel produces values that follow a statistical distribution with mean $\mu_z$ 
and variance $\sigma^2_z$. Shown are schematic probability distributions of 
(a) $\mu_z$ and (b) $\sigma^2_z$, which correspond, respectively, to the true 
bifurcation parameter (light blue) and a ``false'' latent parameter that does 
not correspond to any actual parameter (dark blue).}
\label{fig:ILP}
\end{figure}

The VAE regularization term $R_{z}$ measures: (1) how much information about the 
data is captured by the latent representation ($E_1$), (2) the redundancy or 
dependencies among the different latent variables ($E_2$), and (3) the closeness
of the distribution of each dimension of $z$ to $\mathcal{N}(0,1)$, where the 
distribution parameters $\mu_z$ and $\sigma^2_{z}$ both are uniform across 
different input $x$ ($E_3$). Working together, these three components force the 
VAE to learn a parameter representation with minimal information and independent 
parameters~\cite{Lu2020}. As a result, the hidden parameters can be identified
through the statistics of $\mu_z$ and $\sigma^{2}_{z}$ generated by the VAE 
encoder for each dataset. That is, a faithful latent parameter should have a 
high variance in $\mu_z$ and a low mean of $\sigma^{2}_{z}$, indicating that 
the extracted parameter is precise and informative, as illustrated in 
Fig.~\ref{fig:ILP} with the light colors. In contrast, a parameter collapsing 
to the prior and failing to provide useful information will display a low 
variance in $\mu_z$ and high mean $\sigma^{2}_{z}$, as shown schematically in 
Fig.~\ref{fig:ILP} with dark colors. Intuitively, a high level of variance in 
$\mu_z$ indicates that the latent variables have captured the key information 
about the data that can aid the reconstruction of the time series. However, a 
trivial channel with low variance in $\mu_z$ would yield almost the same $\mu_z$ 
for every input, regardless of variations in their underlying dynamics, and is 
thus uninformative. A low mean $\sigma^{2}_{z}$ ensures that the latent variables 
are not only informative but also precise, thereby reducing the uncertainty. This 
precision is regulated by both the KL divergence term $E_3$ and the mutual 
information term $E_1$, as uncertain latent variables would reduce the overall 
information captured by the VAE.

To demonstrate our unsupervised-learning scheme for anticipating critical 
transitions, we use simulated datasets from the chaotic Lorenz 
system~\cite{lorenz1963} with one or two bifurcation parameters and the 
1D wave system described by the Kuramoto-Sivashinsky 
equation~\cite{Kuramoto:1978,Sivashinsky:1980}. 

\begin{figure} [ht!]
\centering
\includegraphics[width=1\linewidth]{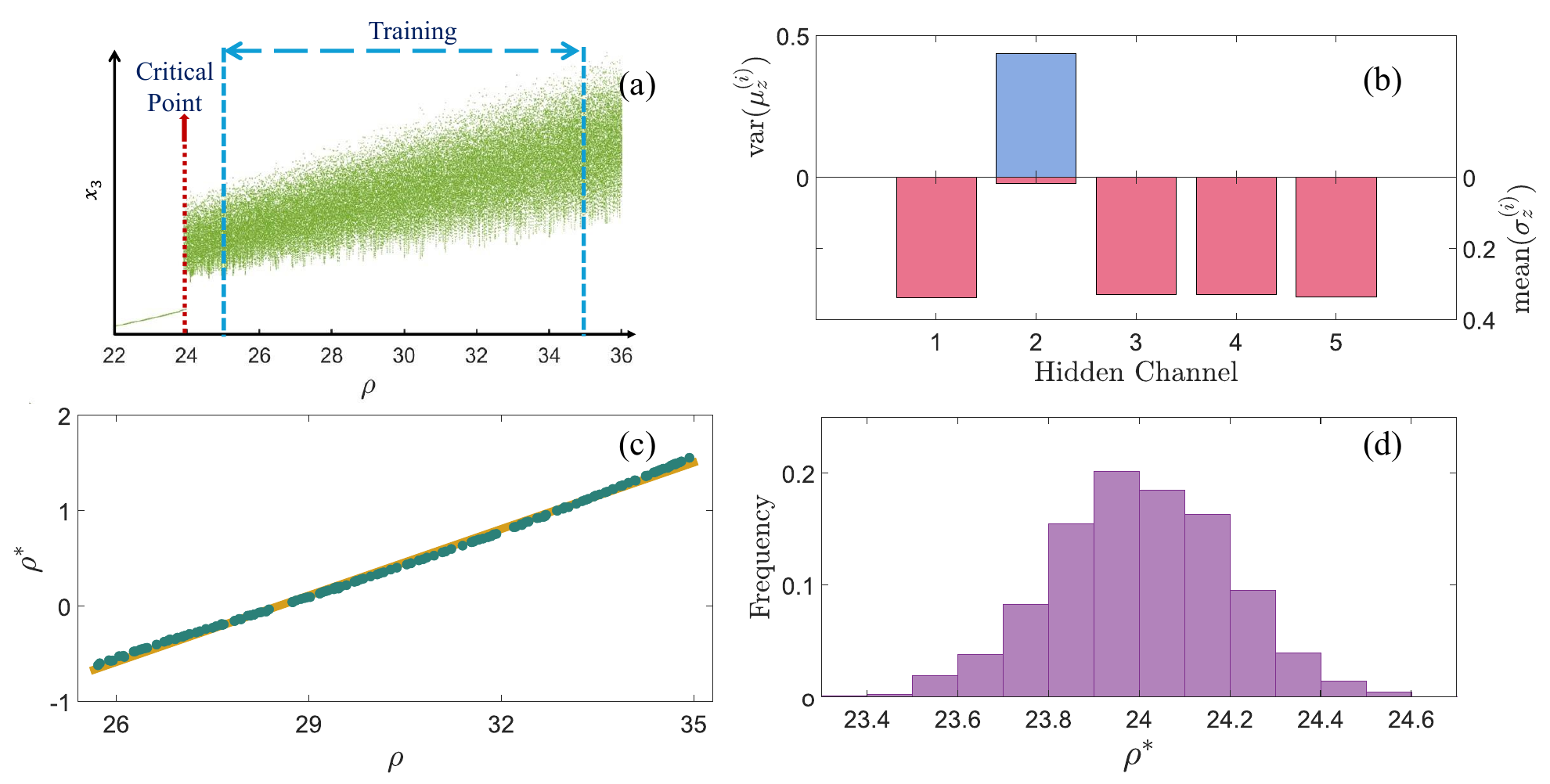}
\caption{Unsupervised-learning based anticipation of a critical transition in 
the chaotic Lorenz system with $\rho$ as the single bifurcation parameter for 
$\beta = 8/3$ and $\sigma = 10$. (a) Bifurcation diagram. The VAE is trained 
with time series from the interval specified by the two vertical blue dashed 
lines. (b) VAE identification of the bifurcation parameter according to the 
behaviors of the variance of the mean $\mu_z$ of the latent parameter (blue) 
and the mean of its variance $\sigma^2_z$ (red) for the five parameter channels 
in the VAE. (c) VAE-detected parameter versus the ground truth physical parameter 
(green dots) and the linear fitting (solid yellow line). (d) Histogram of the 
predicted critical point $\rho^*$ obtained from $1000$ random realizations 
of the reservoir computer. The resulting distribution centers about the ground 
truth $\rho_c \approx 24.06$.}
\label{fig:Lorenz_1d}
\end{figure}

The Lorenz system is given by $\dot{x_1} = \sigma (x_2-x_1)$,
$\dot{x_2} = x_1(\rho -x_3) - x_2$, and $\dot{x_3} = x_1x_2 - \beta x_3$, where 
$\sigma$, $\rho$, and $\beta$ are parameters. We first consider the case where
the two parameters $\sigma$ and $\beta$ are fixed: $\sigma = 10$ and $\beta = 8/3$,
and $\rho$ is the single bifurcation parameter. Figure~\ref{fig:Lorenz_1d}(a) shows 
the bifurcation diagram of the system versus $\rho$. As $\rho$ decreases, a 
critical transition leading to a sudden change in the dynamics of the system 
occurs. For $24 < \rho < 36$, the system functioning is ``normal'' in the sense 
that the system exhibits healthy oscillations. As $\rho$ decreases, a boundary 
crisis~\cite{GOY:1983} occurs at the critical value $\rho_c \approx 23.99$ 
(indicated by the red dotted line), after which the oscillations cede following 
a chaotic transient. 

For training the VAE, we randomly sample the parameter $\rho \in [25 \,\, 35]$, as
indicated by the two blue vertical lines in Fig.~\ref{fig:Lorenz_1d}(a). During 
the training, the VAE uses five latent parameter channels. Since we have only one 
varying bifurcation parameter, the trained VAE should only make use of one latent 
channel, and the rest four channels will collapse to the prior. 
Figure~\ref{fig:Lorenz_1d}(b) shows the statistics of the extracted distribution of
the five latent parameter channels. There is one channel with a high variance in 
$\mu_z$ (blue bins) and a low mean $\sigma^{2}_{z}$ (red bins), which shows that 
the VAE has correctly detected the number of hidden parameters. For this particular
channel, the high variance in $\mu_z$ means that the VAE has captured the driving 
factor that alters the dynamical features in the time-series data. The low mean 
value of the variance $\sigma^{2}_{z}$ implies that the corresponding Gaussian 
distribution from sampling $z$ is almost an impulse function with a narrow width, 
so we have $z\approx\mu_z$. Given these features in the $\mu_z$ and 
$\sigma^{2}_{z}$, the extracted $\mu_z$ can be interpreted as an estimation of 
the identified bifurcation parameter whose variations are responsible for the 
time series $x$ from different values of the bifurcation parameter. For 
performance evaluation, we compare the extracted latent parameter from the VAE 
with the true parameter used to generate the simulated datasets, as shown in 
Fig.~\ref{fig:Lorenz_1d}(c). Remarkably, the detected parameter $\rho^*$ is 
linearly related to the true parameter $\rho$ randomly chosen from the normal 
distribution. This linear relation suggests a one-to-one correspondence between 
the nontrivial latent and the ground truth parameter, as shown by the yellow 
fitting line in Fig.~\ref{fig:Lorenz_1d}(c).

The VAE estimated parameter $\hat{z}=\mu_z$ and the corresponding time series 
data allow us to train the reservoir computer: for each $\hat{z}$ 
value, training is done such that the reservoir machine can predict the state 
evolution of the input data for several Lyapunov times. In the testing phase, we 
apply a parameter change $\Delta \hat{z}$. For each resulting parameter value, we 
test whether the predicted attractor is the ground-truth chaotic attractor, where 
$\Delta \hat{z}$ can be varied systematically for the critical point $\rho^*$ to 
be determined. Figure~\ref{fig:Lorenz_1d}(d) shows a histogram of the predicted 
critical point $\rho^*$, where the linear parameter transformation 
$\rho^* = C_1 \hat{z}  + C_2$ is used with $C_1 = 4.3$ and $C_2 = 28.5$ so as to 
map the $z$ values to the real domain $\rho^*$. Averaging over an ensemble of 
1000 independent random reservoir realizations, the value of the critical point 
lies in the interval $\rho_c^*\approx 24\pm 0.5$, as shown in 
Fig.~\ref{fig:Lorenz_1d}(d). These results show that the parameter-driven 
reservoir computer, with the parameter information provided by the VAE, is capable 
of accurately predicting the crisis.

Note that mapping the final prediction of the transition point back to the 
ground-truth parameter space of $\rho$ is solely for validating the proposed 
method. In an applied scenario where the ground-truth value or even the 
bifurcation parameter is not known, relying on predictions in the latent 
space of $\hat{z}$, as determined by the VAE, is sufficient for both forecasting 
and monitoring, as the VAE-extracted latent variable naturally takes on the role 
of an effective bifurcation parameter.

\begin{figure} [ht!]
\centering
\includegraphics[width=1\linewidth]{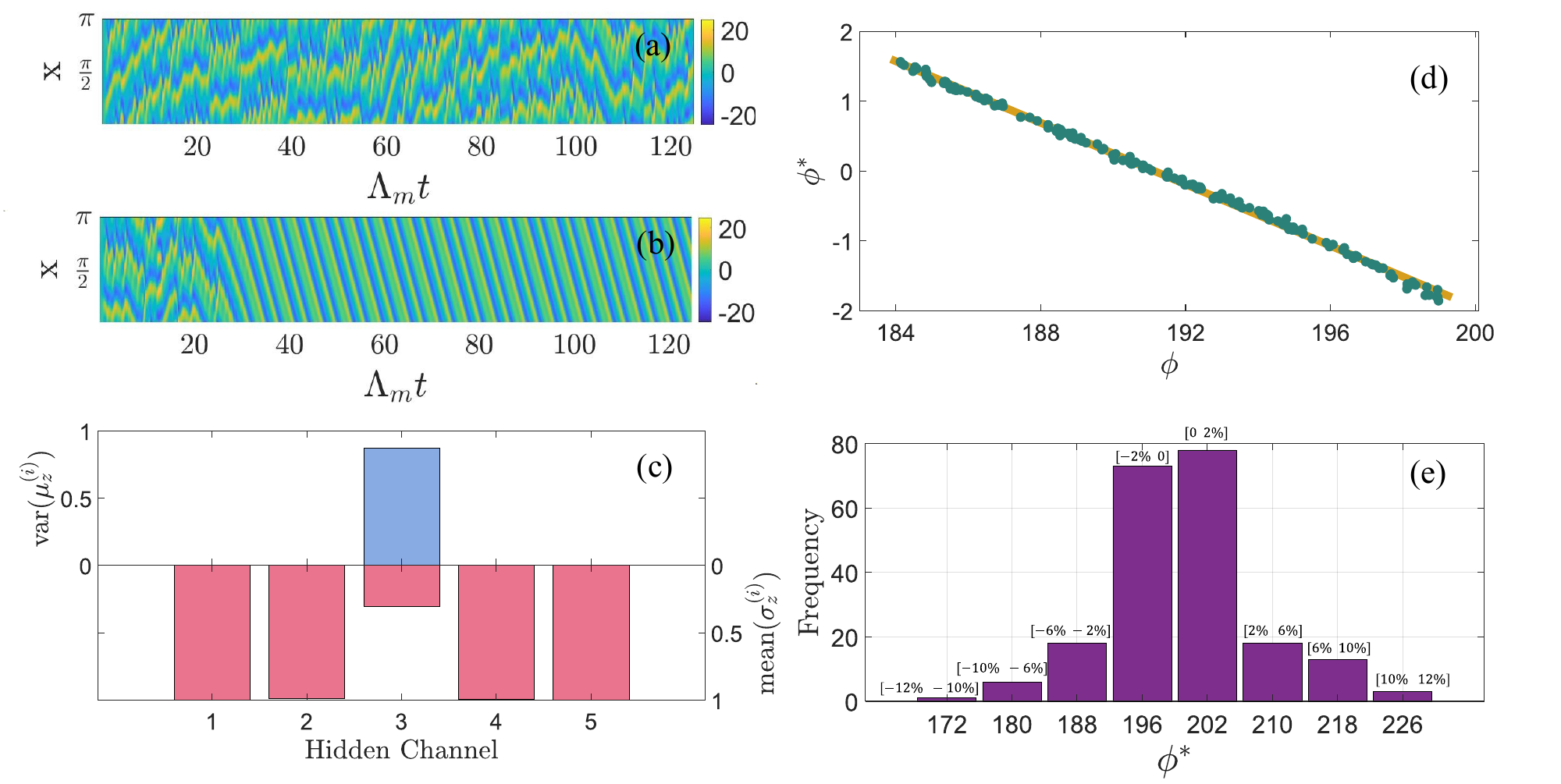}
\caption{Predicting critical transition in the nonlinear wave system described
by the Kuramoto-Sivashinsky equation. Two examples of sustained and transient 
spatiotemporal chaos are shown for (a) $\phi = 199.94 \leq \phi_c$ and 
(b) $\phi = 200.14 \geq \phi_c$. (c) Variance of $\mu_z$ (blue) and mean of 
$\sigma^2_z$ (red) for the five parameter channels in the trained VAE.  
(d) VAE's detected parameters versus the ground truth parameter 
(green dots) and linear fitting (yellow solid line). (e) Histogram of the predicted 
critical point $\phi^*$ and the relative errors with respect to the ground truth 
$\phi_c \approx 200.04$. The distribution is collected from 210 random reservoir
realizations.} 
\label{fig:KS}
\end{figure}

We next study the spatiotemporal system described by the 
Kuramoto-Sivashinsky equation~\cite{Kuramoto:1978,Sivashinsky:1980}:
$\partial_t u = -\nu \partial_x^4 u -\phi(\partial_x^2u+ u\partial_xu)$, where 
$u(t,x)$ is the dynamical variable at position $x$ and time $t$, $\nu = 4$ is 
the damping coefficient and $\phi$ is a bifurcation parameter. The spatial domain
is $0 \leq x \leq \pi$ with the periodic boundary condition. A critical
transition occurs at $\phi_c \approx 200.04$, where there is sustained and 
transient spatiotemporal chaos for $\phi < \phi_c$ and $\phi > \phi_c$, 
respectively. Two examples are shown in Figs.~\ref{fig:KS}(a) and \ref{fig:KS}(b), 
respectively. For training the VAE, we randomly sample $500$ points in 
$\phi \in [184 \,\, 199]$ that belongs to the regime of sustained spatiotemporal 
chaos. During the training, the VAE uses five latent parameter channels. 
Figure~\ref{fig:KS}(c) shows that only one channel exhibits a high variance in 
$\mu_z$ and a low mean $\sigma^{2}_{z}$, indicating that the unsupervised learning 
scheme has correctly identified the hidden parameter. Figure~\ref{fig:KS}(d) shows 
the extracted latent parameters $\hat{z}$ in comparison with the ground truth 
parameters $\phi$ for generating the datasets (represented by the green points). 
The results suggest the existence of a linear transformation mapping the estimated 
parameter $\hat{z}$ to the real values $\phi$, as indicated by the yellow line in 
Fig.~\ref{fig:KS}(d). The detected parameter $z$ along with their corresponding 
time-series data are then used to train the parameter-driven reservoir computer. 
To make predictions, a new control parameter value is generated with a small 
change $\Delta \hat{z}$, and the output of the reservoir computer is evaluated to 
determine whether the predicted state corresponds to sustained or transient chaos. 
Figure~\ref{fig:KS}(e) shows a histogram of the predicted critical point $\phi^*$, 
where the linear transformation $\phi^* = C_1 \hat{z}  + C_2$ with $C_1 = -4.5$ 
and $C_2 = 191.13$ is used to bring back the $z$ to the real parameter $\phi^*$. 
The histogram reveal that about $57\%$ of the predicted values of $\phi^*$ have 
relative error $|\delta| < 2\%$. For the accuracy relaxed to $|\delta| < 10\%$, 
the fraction of correct prediction becomes $98\%$. These results indicate that 
our unsupervised learning framework is applicable to anticipating critical 
transitions in spatiotemporal chaotic systems.

Additional tests have been performed for the Lorenz system with two bifurcation 
parameters [Sec.~S4 in SI~\cite{SI}]. The issue of partial state observation is 
addressed [Sec.~S5 in SI~\cite{SI}] and a chaotic ecosystem violating the sparsity 
condition has also been tested [Sec.~S6 in SI~\cite{SI}]. For all the systems 
tested, the selection of the optimal hyperparameters is described in Sec.~S7 in 
SI~\cite{SI}.

To summarize, our unsupervised learning framework integrates VAE with 
parameter-driven reservoir computing, where VAE is used to identify the key 
bifurcation parameter(s) and its (their) variations. The extracted parametric 
information is fed into a parameter-driven reservoir computer for anticipating 
critical transitions. The framework was tested using diverse examples in different 
settings. In all cases, the robustness and accuracy in anticipating critical
transitions were demonstrated. Our work advocates the use of modern machine 
learning methods in predicting the future behaviors of nonlinear and complex 
dynamical systems in real-world scenarios, with potential applications in 
different fields.

This work was supported by AFOSR under Grant No.~FA9550-21-1-0438 and by ARO
under Grant No.~W911NF-24-2-0228. L.-W.K. was supported by the Eric and Wendy 
Schmidt AI in Science Postdoctoral Fellowship, a Schmidt Futures Program.

\clearpage
\appendix

\section{Previous approaches to anticipating critical transitions and limitations} \label{appendix_a}

Previously, anticipating critical transitions leading to a system collapse was done
by using the sparse optimization approach to finding the governing equations of the 
target dynamical system~\cite{WYLKG:2011} or by exploiting reservoir-computing based
machine learning~\cite{KFGL:2021a}. Both approaches have limitations: the former 
requires that the system equations have a sparse structure and the latter 
necessitates information about the bifurcation parameter whose variations lead 
to a critical transition. The machine-learning approach can thus be categorized as
supervised learning. The main motivation for our work is the real-world situations 
where nothing about the system parameters is known and the only available information 
is some time-series data observed or measured from a number of dynamical variables 
of the system. To anticipate a critical transition in such a case, the parameter 
variations need to be ``learned'' from the data without any ``teacher,'' rendering 
unsupervised the learning process. The main contribution of our work is an 
unsupervised learning framework tailored to anticipating critical transitions 
induced by parameter changes but no prior information about any parameter of the 
system is needed.

\subsection{Sparse optimization for finding the governing equations from data} \label{appendix_a1}

An earlier approach to anticipating critical transitions was
articulated~\cite{WYLKG:2011} in 2011, which is based on finding the governing 
equations of the system from data~\cite{CM:1987,Bollt:2000,YB:2007}. The idea is 
that the equations of certain nonlinear dynamical systems possess a sparse 
structure: they contain a small number of terms expressible in terms of elementary 
mathematical functions, such as power-series or Fourier-series terms. Well studied,
paradigmatic systems in nonlinear dynamics such as the Lorenz 
system~\cite{lorenz1963}, the R\"{o}sller system~\cite{rossler1976equation},
the H\'{e}non map~\cite{Henon:1976}, and the standard map belong to this category. 
For such systems with a ``simple'' equation structure, the problem of finding the 
governing equations becomes that of finding the coefficients associated with these 
terms in a power series or a Fourier series, which can be solved by the standard 
sparse-optimization methods such as compressive 
sensing~\cite{CRT:2006a,CRT:2006b,Donoho:2006,Baraniuk:2007,CW:2008}.
This equation-finding approach was extended to solving a variety of problems in 
nonlinear systems and complex networks~\cite{WLG:2016,Lai:2021} such as unveiling 
the structure of complex oscillator networks~\cite{WYLKH:2011} and social 
networks~\cite{WLGY:2011}, data-driven forecasting of synchronizability of complex 
networks~\cite{SNWL:2012}, detection of hidden nodes~\cite{SWL:2012,SLWD:2014}, and 
reconstruction of propagation or spreading networks~\cite{SWFDL:2014}. 

The sparsity condition required for this approach is in fact self-sabotage: while 
it is the reason that the powerful sparse-optimization methods are applicable, it 
also represents a condition that many real-world systems do not meet. As such, 
this approach is limited to systems whose equation structures are sparse. There 
were also previous works on anticipating a common class of critical transitions 
known as tipping at which a ``normal'' stable steady state of the system is 
destroyed and replaced by another one that can be catastrophic (e.g., population 
extinction in an ecosystem)~\cite{Scheffer:2004,Schefferetal:2009,Scheffer:2010,WH:2010,TC:2014,JHSLGHL:2018,YLTLZCX:2018,JHL:2019,meng2020tipping,MLG:2022} through detecting 
early-warning signals~\cite{SBBBCDHNRS:2009,DG:2010,BH:2012,CLLLA:2012,DVKG:2012,BRH:2013,LeemputEtal:2014,Boers:2018,BSPSLAB:2021}. 
These methods are based on the observation that, as the system approaches a tipping 
point, the statistical fluctuations of the dynamical variables would increase 
dramatically, as the leading eigenvalue associated with the stable steady state 
is about to approach zero.

\subsection{Limitation of previous parameter-driven reservoir computing} \label{appendix_a2}

A limitation of parameter-driven reservoir computing is that it requires the 
knowledge of the bifurcation or control parameter that may change with time.
In particular, during the standard training process~\cite{KFGL:2021a}, the 
bifurcation parameter values need to be injected into the neural network together 
with the state time series of the target system. If the exact parameter values are 
not known, the relative trend of the parameter may also suffice. For instance, 
suppose we have four different training data sequences, such as the population time 
series of some animal species collected from four different locations labeled as 
A, B, C, and D, respectively. Suppose there is one dominant parameter, the average 
temperature, which makes the population dynamics different among the four places. 
Then, to make any interpolation or extrapolation possible, we at least need to 
know the order (descending or ascending) of the parameter among the four data 
sets, be it A-B-C-D or A-C-B-D, for example. With different orders, we could end 
up with qualitatively different results, especially if we perform an extrapolation, 
as we may have the parameter tendency qualitatively wrong. Overcoming this difficulty 
requires identifying uncontrolled dynamical parameters that cause the variations 
in the dynamics of a set of observed trajectories. This ensures 
interpretability~\cite{Zhang2021} without compromising the flexibility inherent 
in machine learning. Simply learning a black-box predictor for each trajectory 
does not provide an accurate extrapolation of the dynamics' changes among 
different trajectories~\cite{Rudin2019}. 

It was demonstrated that the parameters governing the dynamics of a complex 
nonlinear system can be encoded in the learned readout layer of a reservoir
computer~\cite{Alao2021}. Another approach is a variational-autoencoder (VAE) 
architecture consisting of an encoder that extracts the physical parameters 
characterizing the system dynamics and a decoder that acts as a predictive model 
and propagates an initial condition forward in time given the extracted 
parameters~\cite{Lu2020}. It is worth noting that VAEs are widely used for 
dimensionality reduction and unsupervised learning tasks~\cite{Goodfello2016} 
and for studying a wide variety of physical phenomena. 

\section{Variational autoencoders (VAEs)} \label{appendix_c}

\begin{figure} [ht!]
\centering
\includegraphics[width=\linewidth]{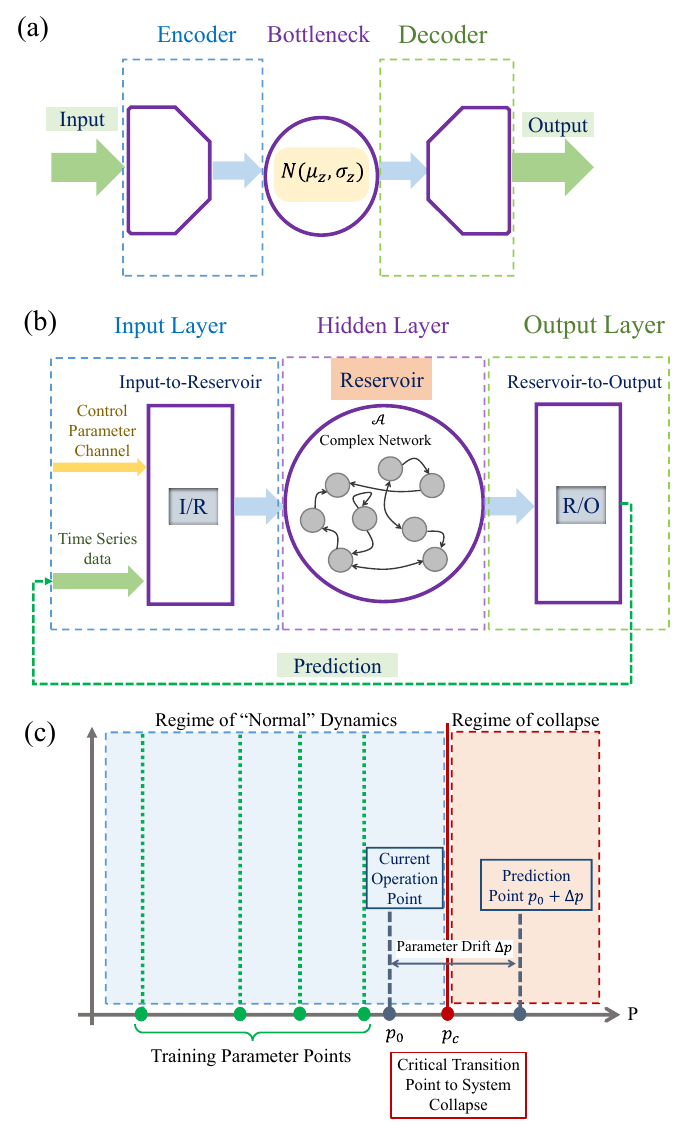}
\caption{VAEs and parameter-driven reservoir computing. (a) Structure of a VAE.
(b) Structure of parameter-driven reservoir computing, where the input consists
of two components: time-series data and the values of the VAE-extracted bifurcation 
parameter. (c) Scheme of training for predicting critical transitions. The light
blue region represents the regime of normal system operation while the light 
orange region denotes the parameter regime of collapse. A critical transition 
from normal operation to collapse occurs at the parameter value $p_c$, and 
$p_0 < p_c$ is the current operation point. Historical time-series data from a 
small number of parameter values in the normal regime are used for training, 
as indicated by the four vertical dashed green lines. Prediction is done for 
$p = p_0 + \Delta p$, where $\Delta p > 0$ is a parameter drift.
}
\label{fig:RC}
\end{figure}

VAEs~\cite{Kingma2013,rezende2014stochastic} are a generative model in 
unsupervised learning with the advantages of learning intricate and continuous 
latent representations of complex datasets. VAEs combine the principles from 
variational inference with deep neural networks to efficiently capture the 
underlying structure of the input data. The core of a VAE architecture is an 
encoder-decoder framework, as shown in Fig.~\ref{fig:RC}(a). The encoder part 
can be regarded as a stochastic process that samples a representation 
$z\sim Q_{\phi}(z|x)$ of the input data $x$ from a distribution $Q_{\phi}(z|x)$ 
parameterized by the functions of the input, where $\phi$ represents the encoder 
parameters. This distribution typically follows a Gaussian form with mean $\mu$ 
and standard deviation $\sigma$. More specifically, the encoder first yields the 
distribution parameters $(\mu,\sigma)=q_{\phi}(x)$ for each input data $x$. Then 
the sampling process is denoted as $z = \mu + \sigma \epsilon$, where $\epsilon$ 
is drawn from a standard normal distribution. The decoder part, with a generative 
process that can be formulated as $\hat{x} \sim P_{\theta}(\hat{x}|z)$, is 
responsible for the task of reconstructing the input data $x$ from the sampled 
latent variable $z$, with $\theta$ representing the decoder parameters. 

The training goal of a VAE is maximizing the evidence lower bound (ELBO), which 
is a variational approximation of the logarithmic-likelihood of the data. The 
ELBO encompasses two key components: a reconstruction term and Kullback-Leibler 
(KL) divergence between the approximate posterior and the prior distributions 
of the latent variables, where the former ensures accurate data reconstruction
and the latter regularizes the learned latent space, preventing it from deviating 
excessively from a predefined prior distribution (often a unit Gaussian).
The training of a VAE is thus a minimax optimization problem - maximizing the 
reconstruction fidelity while minimizing the channel capacity in the bottleneck. 
The overall loss function can be written as 
\begin{equation} \label{eq:b1}
	\mathcal{L} = \frac{1}{N} \sum_{i=1}^{N} \left( x_i - \hat{x}_i \right)^2 + \text{KL} \Big( q_{\phi}(z|x) \parallel p(z) \Big),
\end{equation}
where $x_i$ and $\hat{x}_i$ are the $i^{th}$ data point and the reconstructed
point, respectively. Here, the second term is the regularization term $R_z$. It 
is equivalent to the expression in the main text where $R_z$ is decomposed into 
three terms $R_z=E_1+E_2+E_3$. While the decomposed expression enables an
intuitive understanding, the expression in Eq.~(\ref{eq:b1}) facilitates numerical 
calculation. The optimization of the VAE parameters with respect to this 
loss function can be achieved through backpropagation. Overall, the advantages 
of VAEs are modeling complex data distributions and generating novel samples 
from the learned latent space, making them particularly well-suited for diverse 
applications such as data generation and dimensionality reduction. A special 
feature of VAEs that is particularly relevant to our work is their ability to 
extract informative latent parameters from time-series data. {\em When combined 
with parameter-driven reservoir computing, the framework makes it possible to 
anticipate critical transitions even without direct knowledge about the 
bifurcation parameter, as its variations can be extracted solely from the time 
series data}.

\section{Parameter-driven reservoir computing} \label{appendix_d}

Reservoir computing is a class of recurrent neural 
networks~\cite{Jaeger:2001,MNM:2002,MJ:2013} capable of self dynamical evolution.
The core of reservoir computing is a hidden layer that contains a complex network 
of mutually interacting nonlinear neurons. The architecture consists of an input 
layer, the reservoir network in the hidden layer (the recurrent layer), and an 
output layer. The reservoir network and the input layer are both randomly 
predetermined and fixed throughout the training and testing phases; Only the output 
layer is optimized according to the training data and the loss function, where a 
ridge regression is usually sufficient for the optimization. The training process 
is thus finding a linear combination of the nonlinear response signals generated 
in the reservoir network.

Parameter-driven reservoir computing has an input parameter channel in addition
to the data input channels, as shown in Fig.~\ref{fig:RC}(b), which enables each 
neuron in the network to receive specific parameter values associated with input 
time-series data~\cite{KFGL:2021a}, making the reservoir computer ``aware'' of
the parameter changes in the target system. The training process is illustrated
in Fig.~\ref{fig:RC}(c), where $p$ is the bifurcation parameter of the target 
system. As $p$ varies, a critical transition occurs at $p_c$ that separates
the regime normal system functioning ($p < p_c$, light blue) and collapse after
a transient ($p > p_c$, light orange). Assume that $p$ increases slowly with time
and let $p_0$ be the parameter value of the current system operation. The training 
is done using the historical time series from a small number of distinct parameter 
values, all to the left of $p_0$ in the normal regime, as indicated by the four 
vertical dashed green lines in Fig.~\ref{fig:RC}(c). For each of these parameter 
values, training is done such that the reservoir computer generates accurate 
short-term prediction of the system's state evolution. This adaptive training 
paradigm empowers the reservoir computer with the ability to anticipate critical 
transitions induced by parameter variations. In particular, to predict a potential 
system collapse in the future, we assume a parameter change $\Delta p > 0$ and
let the reservoir computer generate the dynamical behavior at the future parameter
value $p_0 + \Delta p$ through self evolution. For $p_0 + \Delta p < p_c$, a 
well trained reservoir computer should generate the normal attractor of the 
target system. However, for $p_0 + \Delta p > p_c$, the reservoir computer should
generate an attractor that is indicative of system collapse. 

\section{Unsupervised learning for anticipating critical transition of Lorenz system
with two independent bifurcation parameters} \label{appendix_e}

\begin{figure*} [ht!]
\centering
\includegraphics[width=1\linewidth]{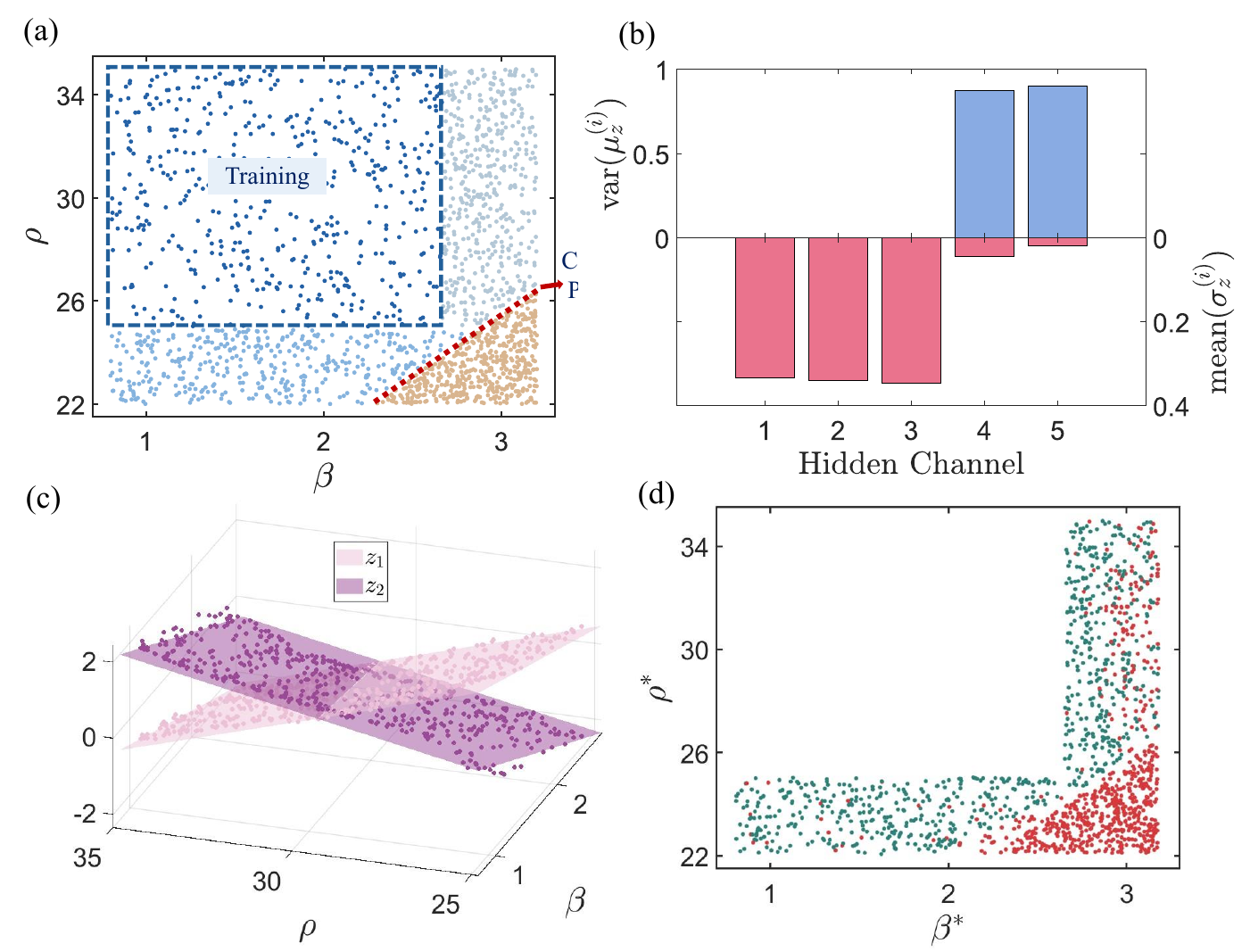}
\caption{Unsupervised learning for anticipating critical transition of Lorenz system 
with two independent bifurcation parameters $\rho$ and $\beta$. (a) 2D bifurcation 
diagram with respect to the parameter plane ($\rho$, $\beta$) for $\sigma = 10$. The 
VAE is trained with time series taken within the blue regime in which the system 
operates normally. The training parameter values is indicated by the light blue 
dashed lines. (b) Identification of the bifurcation parameters from the variance 
of the mean $\mu_z$ (blue) and the mean of the variance $\sigma^2_z$ (red) for the 
five parameter channels in the VAE. (c) VAE's detected parameter values in comparison 
with the ground truth physical parameters, where the dark and light purple dots 
corresponding to the first and second extracted latent parameter, respectively. Each 
shaded surface represents a linear combination of the ground truth parameters (dark 
and light purple). (d) Predicted parameter points transformed into the real domain. 
The red dots represent points with unhealthy behaviors where the reservoir computer 
output exhibits a critical transition, while the green red dots denote healthy 
behaviors - those without any critical transition.}
\label{fig:Lorenz_2d}
\end{figure*}

For a real-world system, the presence of multiple bifurcation parameters can be
expected. Can our unsupervised learning framework be effective for anticipating
a critical transition in such a case? As a concrete example, we consider the 
Lorenz system with $\rho$ and $\beta$ as the two independent bifurcation parameters.

Figure~\ref{fig:Lorenz_2d}(a) shows a 2D bifurcation diagram of the system with
respect to the parameters $\rho$ and $\beta$. For parameters in the blue region, the 
system functioning is ``normal'' in the sense that its dynamical behavior 
exhibits ``healthy'' oscillations. The points belonging to the yellow region 
correspond to an abnormal behavior of the system. These two regions are separated 
by a critical curve (a one-dimensional set of critical points) as indicated by the red 
dotted line. Near these points, the variation of parameters leads to a sudden change 
in the dynamics triggered by a boundary crisis~\cite{GOY:1983}.

For training the VAE we randomly sample $500$ points with $\rho\in [25 \,\, 35]$ and 
$\beta \in [0.8 \,\,\, 2.66]$, as indicated by the blue rectangle in 
Fig.~\ref{fig:Lorenz_2d}(a). Similar to the case of a single bifurcation parameter,
during the training, the VAE uses five latent parameter channels. The number of 
hidden parameters is determined using the statistics of the extracted distribution 
parameters $\mu_z$ and $\sigma^{2}_{z}$ for each dataset. In particular, 
Fig.~\ref{fig:Lorenz_2d}(b) shows the statistics of the extracted distribution 
of the five latent parameter channels. There are two channels with a high variance 
in $\mu_z$ and a low mean $\sigma^{2}_{z}$, indicating that the unsupervised learning
framework has correctly detected the number of hidden parameters. To evaluate the 
method's performance, we plot the extracted latent parameters ($z_1$ and $z_2$) from 
the learning scheme (shown as points) against the true physical parameters ($\rho$ 
and $\beta$) used for generating the simulated datasets, as depicted in 
Fig.~\ref{fig:Lorenz_2d}(c), where dark (light) purple points correspond to the first 
(second) latent parameter. The results indicate that the latent parameters can be 
interpreted as a linear combination of the ground truth parameters, where the shaded 
surfaces represent the results of the linear combination of the actual ground truth 
parameters.

Using the VAE's detected parameter and their corresponding time series data, we train 
the parameter-driven reservoir computer. For each pair of the detected parameters, 
training is done such that the reservoir machine can predict the state evolution of 
the input data for several Lyapunov times. After the training, we apply parameter 
changes $\Delta z_1$ and $\Delta z_2$ in a systematic fashion by dividing the 
parameter plane into two sets of potential testing points. The first set comprises 
points with healthy behavior but lying outside the training rectangle, as depicted 
by the light blue points in Fig~\ref{fig:Lorenz_2d}(a). The second set consists of 
points from the yellow region in Fig~\ref{fig:Lorenz_2d}(a), which generate an abnormal 
behavior in the sense of lack of oscillations in the long-term dynamics. For each 
resulting parameter value, we assess whether the predicted system state represents 
a chaotic attractor. The outcomes of one iteration of the reservoir computer are 
shown in Fig.~\ref{fig:Lorenz_2d}(d), where the green and red dots represent the
predictions of the healthy and unhealthy behavior, respectively. The red (green) 
points within the first (second) set indicate an unhealthy (healthy) behavior wrongly 
labeled as healthy (unhealthy). For illustrative purposes, we apply a linear 
transformation: 
\begin{align} \nonumber
[\rho^* \,\, \beta^*]^T= C [z_1, z_2, 1]^T, 
\end{align}
where
\begin{align} \nonumber
C = \begin{pmatrix}1.74 & -2.45 & 30.15\\-0.48 & -0.33 & 1.72\end{pmatrix},
\end{align}
to bring the detected parameters $z_1$ and $z_2$ to their real values $\rho^*$ 
and $\beta^*$. To assess the prediction performance, we compute the false positive 
rate (FPR) and a false negative rate (FNR), where the former represents the fraction
of the red points in the first set and the latter is the fraction of the green points 
in the second set. We obtain $\mbox{FPR} \approx 23 \%$ and $\mbox{FNR} \approx 1 \%$. 
A low FNR is desirable because it indicates a small miss rate, suggesting that our
unsupervised learning framework is capable of predicting the critical bifurcation 
with small errors even with two independent bifurcation parameters.

\section{Anticipating critical transition with partial state observation} \label{appendix_f}

The results so far presented are obtained under the assumption that the bifurcation
parameter of the target system is inaccessible but the time series from all the 
dynamical variables of the target system can be obtained, corresponding to full 
state observation. In real applications, it may occur that only a subset of the 
dynamical variables can be observed or measured, a situation referred to as partial 
state observation. Is our unsupervised learning framework still able to predict 
critical transitions?

\begin{figure} [ht!]
\centering
\includegraphics[width=1\linewidth]{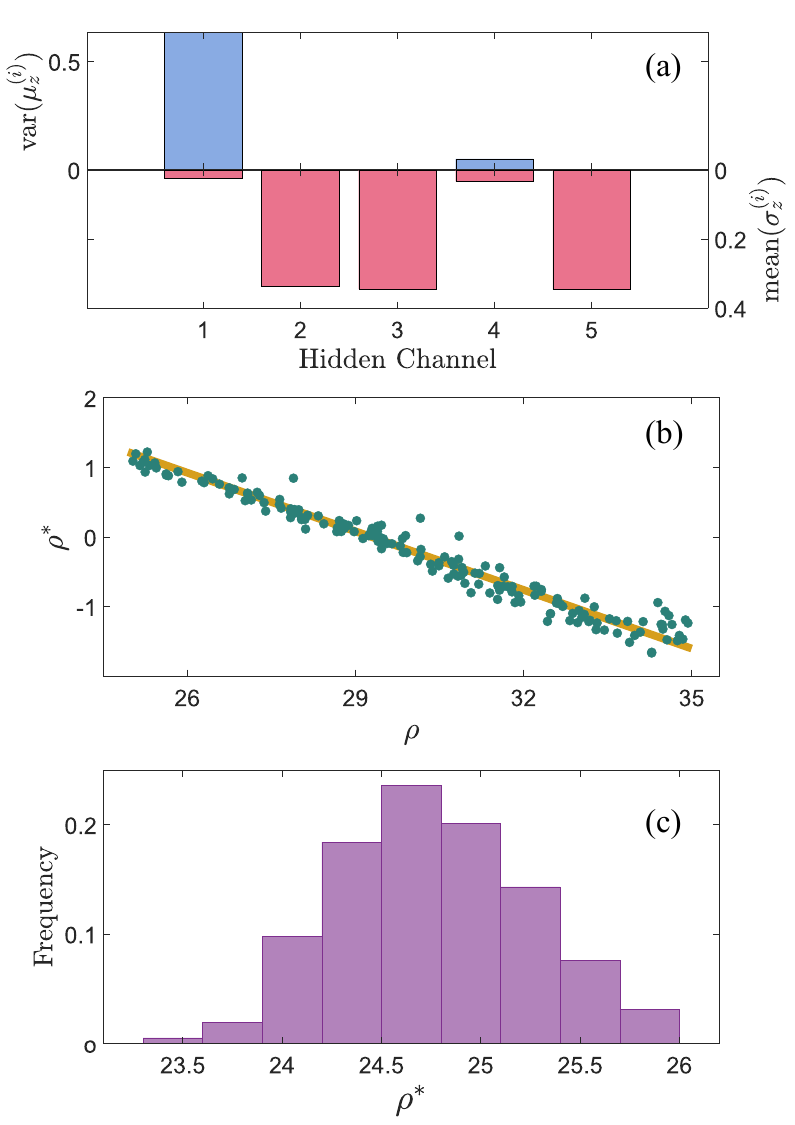}
\caption{Unsupervised-learning based prediction  critical transition in the Lorenz 
system with partial state observation. (a) Variance of $\mu_z$ (blue) 
and mean of $\sigma^2_z$ (red) for the five parameter channels in the VAE. (b) 
VAE's detected parameters versus the ground truth physical parameters (green dots) 
and the linear fitted line (solid yellow line). (c) Histogram of the predicted 
critical point $\rho^*$, with $1000$ random realizations of the reservoir computer.}
\label{fig:partial}
\end{figure}

In nonlinear dynamics, the well-established methodology to deal with partial state
observation is the classical Takens' delayed-coordinate embedding 
framework~\cite{takens2006detecting}, where the phase space of the system of interest
can be reconstructed even with a single time series. We follow this principle by 
changing the structure of the decoder. Recall that, in the case of full state 
observation, the decoder takes only one time step of state $\hat{x}_i$ to predict 
the next $\hat{x}_{i+1}$. With partial state observation, we apply time-delay 
embedding in the input of the decoder to perform the reconstruction/prediction task. 
Specifically, the input to the decoder now contains several steps of the (partial) 
state variables $\{ \hat{x}_{i-d},\hat{x}_{i-d+1},...,\hat{x}_{i-1},\hat{x}_i\}$. 
The output of the decoder is still the next-step prediction $\hat{x}_{i+1}$.

To give a concrete example, we again use the chaotic Lorenz system,
assuming that only the first variable ($x$) can be measured and the single bifurcation
parameter is $\rho$. Figure~\ref{fig:partial}(a) shows the statistical 
characteristics of the extracted distribution across five latent parameter channels.
Notably, there is only one channel with a substantial variance in $\mu_z$ alongside a 
minimal mean $\sigma^{2}_{z}$, indicating that an accurate identification of the number 
of hidden parameters has been achieved. Figure~\ref{fig:partial}(b) shows the extracted 
parameter values in comparison the true physical parameter values (green dots). A 
comparison between the result in Fig.~\ref{fig:partial}(b) with that in 
Fig.~\ref{fig:Lorenz_1d}(c) for the full-state observation case indicates that, while 
there is information loss due to having access to only partial state observation, it 
is still possible to find a linear transformation (the solid yellow line) to map the 
detected latent parameter $z$ to the true parameter $\rho$. 

We can now train the reservoir computer using the detected parameter $\hat{z}$ and their 
corresponding time series data. Systematically applying small parameter change 
$\Delta \hat{z}$ to check if the long-term dynamics are generated by a chaotic attractor, 
we obtain the critical point $\rho^*$, as shown by the histogram in 
Fig.~\ref{fig:partial}(c), where the linear transformation $\rho^* = C_1 \hat{z}  + C_2$ 
with $C_1 = -3.5$ and $C_2 = 29.3$ is used to bring back the $\hat{z}$ to real-domain 
$\rho^*$. Comparing the histogram with that in Fig.~\ref{fig:Lorenz_1d}(d) for the 
case of full state observation reveals that, despite the adverse effects of 
information loss associated with partial state observation, our unsupervised learning 
framework remains capable of predicting the critical transition. 

\section{Anticipating critical transition in a chaotic ecosystem that violates the sparsity condition} \label{appendix_g}

\begin{figure*} [ht!]
\centering
\includegraphics[width=1\linewidth]{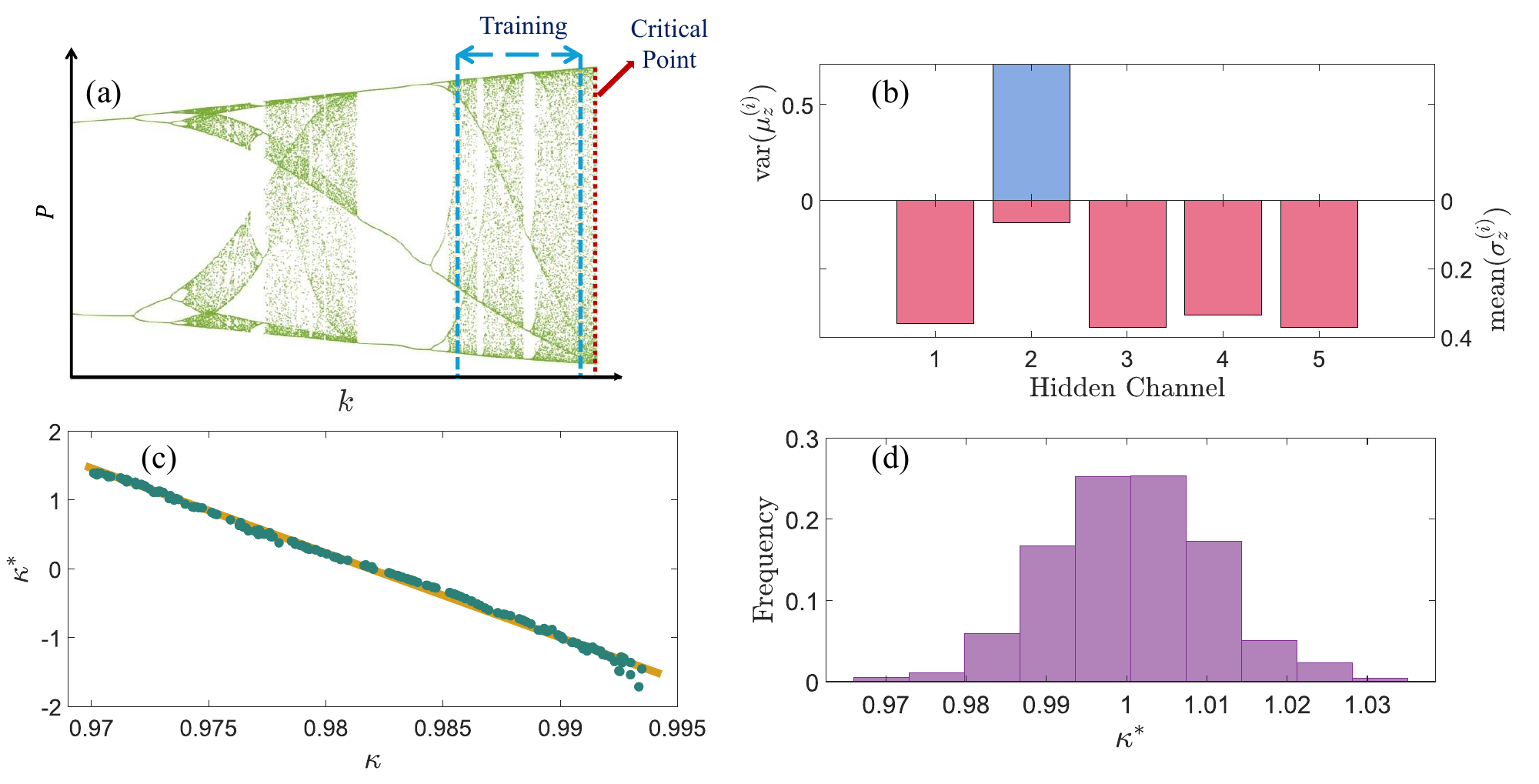}
\caption{Anticipating a critical transition in the chaotic food-chain 
system~\eqref{Eq:foodchain}. (a) Bifurcation diagram with $\kappa$. The fixed values 
of the other parameters are: $x_c = 0.4$, $y_c = 2.009$, $x_p = 0.08$, $y_p = 2.876$, 
$R_0 = 0.16129$, and $C_0 = 0.5$. The VAE is trained with time series taken from the 
blue interval in the green regime in which there is healthy coexistence of all three
species. The training interval is indicated by the two vertical blue dashed lines. 
(b) Variance of $\mu_z$ (blue) and mean of $\sigma^2_z$ (red) for the five parameter 
channels in the VAE. (c) VAE's detected parameters versus the ground truth physical 
parameters (green dots) and the linear fitted line (solid yellow line). (d) Histogram 
of the predicted critical point from $1000$ random realizations of the reservoir 
machine. The distribution centers around the ground-truth critical point $K_c=1.00$.}
\label{fig:FC}
\end{figure*}

The examples treated in the main text belong to the category of dynamical systems 
with a sparse structure. Here we consider an example in ecosystems in which the 
sparsity condition is violated in the sense that the power-series expansions of the 
component functions of the velocity field contain an infinite number of terms, and 
demonstrate that a typical critical transition can be faithfully anticipated by our 
unsupervised learning framework.

There is significant concern about human-activity induced environmental changes
and how these changes may lead to catastrophic event such as sudden species extinction. 
In a typical ecological system, there are two coexisting states: one associated with 
healthy survival and the other with extinction. Changes in the environment can change 
the parameters of the system, triggering a critical shift that occurs when the system 
reaches a critical point, resulting in the disappearance of the survival state and 
leaving only extinction as the final state. To demonstrate that our unsupervised
learning framework can anticipate critical transitions in ecosystem, we consider
the chaotic food chain system of three species~\cite{McCann1994}:
\begin{align}\label{Eq:foodchain}
\begin{split}
    \dot{R} &= R \Big(1-\frac{R}{\kappa} \Big)-\frac{x_cy_c CR}{R+R_0},\\
    \dot{C} &= x_c C \Big( \frac{y_cy_c R}{R+R_0}-1 \Big) - -\frac{x_py_p PC}{C+C_0},\\
    \dot{P} &= x_p P \Big(\frac{y_p C}{C+C_0} -1 \Big),
    \end{split}
\end{align}
where the dynamical variables $R$, $C$, and $P$ are the resource, consumer, and 
predator species density, respectively, $\kappa$ is a parameter characterizing the 
environmental capacity of the resource species and is a bifurcation parameter 
reflecting the effects of environment changes on the system. Other parameters of the 
system ($x_c, y_c, x_p, y_p, R_0$, and $C_0$) are fixed. Figure~\ref{fig:FC}(a) 
shows the bifurcation diagram of the system~\eqref{Eq:foodchain}.

The procedures for generating training data closely resemble those for the chaotic 
Lorenz system. Here, we randomly sample the parameter 
$\kappa \in [0.97 \,\, 0.994]$, as indicated by the two blue vertical lines in 
Fig.~\ref{fig:FC}(a). For each parameter value, we integrate Eq.~\eqref{Eq:foodchain} 
to generate the training time series. Figure~\ref{fig:FC}(b) shows the statistical 
characteristics of the extracted distribution across five latent parameter channels.
It can be seen that there is only one channel with a large variance in $\mu_z$ and 
low mean $\sigma^{2}_{z}$. Figure~\ref{fig:FC}(c) shows the VAE extracted latent 
parameters in comparison with the true physical parameter (green dots). There is a 
linear transformation to map the detected parameter $z$ to the true parameter 
$\kappa$ (yellow solid line). The parameter-driven reservoir computer is trained
using the detected parameter $z$ and their corresponding time series data. To find a 
possible critical transition, a small parameter change $\Delta z$ is applied to the 
trained reservoir computer to check if it produces a survival chaotic attractor.
Figure~\ref{fig:FC}(d) shows a histogram of the predicted critical point $\kappa^*$,
where the linear transformation $\kappa^* = C_1 \hat{z}  + C_2$ is used to bring back the 
extracted $z$ to the real $\kappa^*$. The result in Fig.~\ref{fig:FC}(d) indicates
that our unsupervised learning framework can accurately predict the critical 
transition in the chaotic food-chain system.

\begin{table*}[ht!]
\centering
\caption{Hyperparameter configurations for each example in the main text: Sets A-E 
correspond to, respectively, chaotic Lorenz system with a single control parameter, 
chaotic Lorenz system with two control parameters, chaotic Lorenz system with partial 
state observation, spatiotemporal Kuramoto-Sivashinsky system, and the chaotic food 
chain system.}
\label{tab:T_lorenz_1d}
\begin{tabular}{lc c c c c c}
\hline
\multicolumn{2}{l}{Hyperparameters}                  & Set A             & Set B             & Set C             & Set D             & Set E                                  \\ \hline 
\multicolumn{1}{|l|}{\multirow{3}{*}{VAE}}  & $n_b$    & $100$             & $100$             & $100$             & $50$              & \multicolumn{1}{c|}{$100$}             \\  
\multicolumn{1}{|l|}{}                      & $lr$     & $2\times 10^{-3}$ & $2\times 10^{-3}$ & $2\times 10^{-3}$ & $1\times 10^{-3}$ & \multicolumn{1}{c|}{$5\times 10^{-3}$} \\  
\multicolumn{1}{|l|}{}                      & $\beta_{\text{VAE}}$  & $1\times 10^{-3}$ & $1\times 10^{-3}$ & $1\times 10^{-3}$ & $1$               & \multicolumn{1}{c|}{$1\times 10^{-3}$} \\ \hline \hline
\multicolumn{1}{|l|}{\multirow{7}{*}{}} & $n_r$    & $800$             & $800$             & $800$             & $4000$            & \multicolumn{1}{c|}{$900$}             \\  
\multicolumn{1}{|l|}{Parameter      }                      & $d$      & $550$             & $730$             & $580$             & $1902$            & \multicolumn{1}{c|}{$4$}               \\  
	\multicolumn{1}{|l|}{Driven}                      & $\lambda$   & $1.34$            & $0.55$            & $0.63$            & $0.85$            & \multicolumn{1}{c|}{$2.3$}             \\  
	\multicolumn{1}{|l|}{Reservoir}                      & $k_{in}$ & $0.04$            & $0.08$            & $0.09$            & $0.01$            & \multicolumn{1}{c|}{$3.6$}             \\ 
	\multicolumn{1}{|l|}{Computing}                      & $k_b$    & $3$               & $0.23$            & $1.6$             & 2.78\$            & \multicolumn{1}{c|}{$0.5$}             \\ 
\multicolumn{1}{|l|}{}                      & $b_0$    & $3.5$             & $-1.9$            & $4.4$             & $1.31$            & \multicolumn{1}{c|}{$-2.2$}            \\  
\multicolumn{1}{|l|}{}                      & $\alpha$ & $0.7$             & $0.15$            & $0.17$            & $0.91$            & \multicolumn{1}{c|}{$0.3$}             \\ \hline \hline
\end{tabular}
\end{table*}

\section{Hyperparameter Optimization} \label{appendix_h}

In machine learning, the hyperparameters are those defining the neural-network 
architecture and those associated with the training process, which play an important 
role in achieving accurate testing and generalization results. Determining the optimal 
values of the hyperparameters is computationally demanding and is often implemented as 
a black-box optimization problem. Earlier methods included grid search based on an
exhaustive evaluation of all possible hyperparameter combinations within a constrained 
search space~\cite{NIPS2011}, which are infeasible for high-dimensional problems due to 
the computational cost and lack of scalability. Random search offers a more efficient 
alternative but still suffers from high computational expenses and lacks 
adaptability~\cite{Bergstra2012}. Reinforcement learning has been exploited for finding 
the appropriate architecture of CNNs and for determining training 
parameters~\cite{Zoph2016}, but a deficiency is the absence of a global exploration 
strategy. Bayesian methods, while sophisticated, provide an appealing alternative by 
constructing a probabilistic model over an objective function, utilizing techniques 
such as Gaussian processes or random forests~\cite{williams2006,Breiman2001}. We 
employ the Bayesian optimization method (implemented using the {\it skopt}package 
in Python), for both the VAE and parameter-driven reservoir computing.

A VAE, similar to any deep neural network, typically contains various hyperparameters 
including the number of hidden layers, choice of activation function, optimizer 
selection, batch size ($n_b$), learning rate ($lr$), and the regularization parameter 
($\beta_{\text{VAE}}$). In our work, the number of hidden layers is fixed at 32,
the activation function is chosen to be the ReLU, and the Adam optimization algorithm 
is used. The remaining hyperparameters that need to be optimized are the batch size, 
learning rate, and the regularization parameter. It is worth noting that for the
task of parameter identification, the regularization parameter is particularly
important, as it can affect the extraction of parameters. An optimal choice of the 
regularization parameter can facilitate the learning of disentangled representations. 
Finding an optimized $\beta_{\text{VAE}}$ value, however, demands a delicate balance, 
as excessively high values yield irrelevant parameters while overly low values will 
compromise independence enforcement.

For parameter-driven reservoir computing, the hyperparameters~\cite{KFGL:2021a} 
consist of the predefined parameters characterizing the input architecture and the 
neural network in the hidden layer, which include the network size ($n_r$), the 
average degree ($d$), the spectral radius ($\lambda$), the scaling factor of the 
elements of the data input matrix ($k_{in}$), the parameters associated with the 
parameter-input matrix, and the leakage parameter ($\alpha$). The optimization 
strategy entails training multiple reservoir computers for each hyperparameter set, 
computing the average validation root-mean-square error, and integrating this feedback 
into the Bayesian algorithm. Iterative training with various reservoir realizations 
mitigates the error fluctuations. With a few hundred Bayesian iterations, 
hyperparameter values yielding the lowest validation root-mean-square error across 
all iterations are selected, prioritizing the overall performance over specific 
iteration outcomes. The results of hyperparameter optimization for the examples
in our work are listed in Table~\ref{tab:T_lorenz_1d}.


\begin{thebibliography}{82}%
\makeatletter
\providecommand \@ifxundefined [1]{%
 \@ifx{#1\undefined}
}%
\providecommand \@ifnum [1]{%
 \ifnum #1\expandafter \@firstoftwo
 \else \expandafter \@secondoftwo
 \fi
}%
\providecommand \@ifx [1]{%
 \ifx #1\expandafter \@firstoftwo
 \else \expandafter \@secondoftwo
 \fi
}%
\providecommand \natexlab [1]{#1}%
\providecommand \enquote  [1]{``#1''}%
\providecommand \bibnamefont  [1]{#1}%
\providecommand \bibfnamefont [1]{#1}%
\providecommand \citenamefont [1]{#1}%
\providecommand \href@noop [0]{\@secondoftwo}%
\providecommand \href [0]{\begingroup \@sanitize@url \@href}%
\providecommand \@href[1]{\@@startlink{#1}\@@href}%
\providecommand \@@href[1]{\endgroup#1\@@endlink}%
\providecommand \@sanitize@url [0]{\catcode `\\12\catcode `\$12\catcode
  `\&12\catcode `\#12\catcode `\^12\catcode `\_12\catcode `\%12\relax}%
\providecommand \@@startlink[1]{}%
\providecommand \@@endlink[0]{}%
\providecommand \url  [0]{\begingroup\@sanitize@url \@url }%
\providecommand \@url [1]{\endgroup\@href {#1}{\urlprefix }}%
\providecommand \urlprefix  [0]{URL }%
\providecommand \Eprint [0]{\href }%
\providecommand \doibase [0]{https://doi.org/}%
\providecommand \selectlanguage [0]{\@gobble}%
\providecommand \bibinfo  [0]{\@secondoftwo}%
\providecommand \bibfield  [0]{\@secondoftwo}%
\providecommand \translation [1]{[#1]}%
\providecommand \BibitemOpen [0]{}%
\providecommand \bibitemStop [0]{}%
\providecommand \bibitemNoStop [0]{.\EOS\space}%
\providecommand \EOS [0]{\spacefactor3000\relax}%
\providecommand \BibitemShut  [1]{\csname bibitem#1\endcsname}%
\let\auto@bib@innerbib\@empty
\bibitem [{\citenamefont {Hastings}\ \emph {et~al.}(2018)\citenamefont
  {Hastings}, \citenamefont {Abbott}, \citenamefont {Cuddington}, \citenamefont
  {Francis}, \citenamefont {Gellner}, \citenamefont {Lai}, \citenamefont
  {Morozov}, \citenamefont {Petrovskii}, \citenamefont {Scranton},\ and\
  \citenamefont {Zeeman}}]{hastings2018transient}%
  \BibitemOpen
  \bibfield  {author} {\bibinfo {author} {\bibfnamefont {A.}~\bibnamefont
  {Hastings}}, \bibinfo {author} {\bibfnamefont {K.~C.}\ \bibnamefont
  {Abbott}}, \bibinfo {author} {\bibfnamefont {K.}~\bibnamefont {Cuddington}},
  \bibinfo {author} {\bibfnamefont {T.}~\bibnamefont {Francis}}, \bibinfo
  {author} {\bibfnamefont {G.}~\bibnamefont {Gellner}}, \bibinfo {author}
  {\bibfnamefont {Y.-C.}\ \bibnamefont {Lai}}, \bibinfo {author} {\bibfnamefont
  {A.}~\bibnamefont {Morozov}}, \bibinfo {author} {\bibfnamefont
  {S.}~\bibnamefont {Petrovskii}}, \bibinfo {author} {\bibfnamefont
  {K.}~\bibnamefont {Scranton}},\ and\ \bibinfo {author} {\bibfnamefont
  {M.~L.}\ \bibnamefont {Zeeman}},\ }\bibfield  {title} {\bibinfo {title}
  {Transient phenomena in ecology},\ }\href@noop {} {\bibfield  {journal}
  {\bibinfo  {journal} {Science}\ }\textbf {\bibinfo {volume} {361}},\ \bibinfo
  {pages} {eaat6412} (\bibinfo {year} {2018})}\BibitemShut {NoStop}%
\bibitem [{\citenamefont {Lenton}\ \emph {et~al.}(2019)\citenamefont {Lenton},
  \citenamefont {Rockstr{\"o}m}, \citenamefont {Gaffney}, \citenamefont
  {Rahmstorf}, \citenamefont {Richardson}, \citenamefont {Steffen},\ and\
  \citenamefont {Schellnhuber}}]{lenton2019climate}%
  \BibitemOpen
  \bibfield  {author} {\bibinfo {author} {\bibfnamefont {T.~M.}\ \bibnamefont
  {Lenton}}, \bibinfo {author} {\bibfnamefont {J.}~\bibnamefont
  {Rockstr{\"o}m}}, \bibinfo {author} {\bibfnamefont {O.}~\bibnamefont
  {Gaffney}}, \bibinfo {author} {\bibfnamefont {S.}~\bibnamefont {Rahmstorf}},
  \bibinfo {author} {\bibfnamefont {K.}~\bibnamefont {Richardson}}, \bibinfo
  {author} {\bibfnamefont {W.}~\bibnamefont {Steffen}},\ and\ \bibinfo {author}
  {\bibfnamefont {H.~J.}\ \bibnamefont {Schellnhuber}},\ }\bibfield  {title}
  {\bibinfo {title} {Climate tipping points—too risky to bet against},\
  }\href@noop {} {\bibfield  {journal} {\bibinfo  {journal} {Nature}\ }\textbf
  {\bibinfo {volume} {575}},\ \bibinfo {pages} {592} (\bibinfo {year}
  {2019})}\BibitemShut {NoStop}%
\bibitem [{\citenamefont {Dobson}\ and\ \citenamefont
  {Chiang}(1989)}]{DC:1989}%
  \BibitemOpen
  \bibfield  {author} {\bibinfo {author} {\bibfnamefont {I.}~\bibnamefont
  {Dobson}}\ and\ \bibinfo {author} {\bibfnamefont {H.-D.}\ \bibnamefont
  {Chiang}},\ }\bibfield  {title} {\bibinfo {title} {Towards a theory of
  voltage collapse in electric power systems},\ }\href@noop {} {\bibfield
  {journal} {\bibinfo  {journal} {Sys. Cont. Lett.}\ }\textbf {\bibinfo
  {volume} {13}},\ \bibinfo {pages} {253} (\bibinfo {year} {1989})}\BibitemShut
  {NoStop}%
\bibitem [{\citenamefont {Scheffer}(2020)}]{Scheffer2020}%
  \BibitemOpen
  \bibfield  {author} {\bibinfo {author} {\bibfnamefont {M.}~\bibnamefont
  {Scheffer}},\ }\href {https://books.google.com/books?id=okT_DwAAQBAJ} {\emph
  {\bibinfo {title} {Critical Transitions in Nature and Society}}},\ Princeton
  Studies in Complexity\ (\bibinfo  {publisher} {Princeton University Press},\
  \bibinfo {year} {2020})\BibitemShut {NoStop}%
\bibitem [{\citenamefont {O’Regan}\ \emph {et~al.}(2020)\citenamefont
  {O’Regan}, \citenamefont {O’Dea}, \citenamefont {Rohani},\ and\
  \citenamefont {Drake}}]{o2020transient}%
  \BibitemOpen
  \bibfield  {author} {\bibinfo {author} {\bibfnamefont {S.~M.}\ \bibnamefont
  {O’Regan}}, \bibinfo {author} {\bibfnamefont {E.~B.}\ \bibnamefont
  {O’Dea}}, \bibinfo {author} {\bibfnamefont {P.}~\bibnamefont {Rohani}},\
  and\ \bibinfo {author} {\bibfnamefont {J.~M.}\ \bibnamefont {Drake}},\
  }\bibfield  {title} {\bibinfo {title} {Transient indicators of tipping points
  in infectious diseases},\ }\href@noop {} {\bibfield  {journal} {\bibinfo
  {journal} {J. R. Soc. Interface.}\ }\textbf {\bibinfo {volume} {17}},\
  \bibinfo {pages} {20200094} (\bibinfo {year} {2020})}\BibitemShut {NoStop}%
\bibitem [{\citenamefont {van Nes}\ \emph {et~al.}(2016)\citenamefont {van
  Nes}, \citenamefont {Arani}, \citenamefont {Staal}, \citenamefont {van~der
  Bolt}, \citenamefont {Flores}, \citenamefont {Bathiany},\ and\ \citenamefont
  {Scheffer}}]{NASBFBS:2016}%
  \BibitemOpen
  \bibfield  {author} {\bibinfo {author} {\bibfnamefont {E.~H.}\ \bibnamefont
  {van Nes}}, \bibinfo {author} {\bibfnamefont {B.~M.}\ \bibnamefont {Arani}},
  \bibinfo {author} {\bibfnamefont {A.}~\bibnamefont {Staal}}, \bibinfo
  {author} {\bibfnamefont {B.}~\bibnamefont {van~der Bolt}}, \bibinfo {author}
  {\bibfnamefont {B.~M.}\ \bibnamefont {Flores}}, \bibinfo {author}
  {\bibfnamefont {S.}~\bibnamefont {Bathiany}},\ and\ \bibinfo {author}
  {\bibfnamefont {M.}~\bibnamefont {Scheffer}},\ }\bibfield  {title} {\bibinfo
  {title} {What do you mean, ‘tipping point’?},\ }\href
  {https://doi.org/10.1016/j.tree.2016.09.011} {\bibfield  {journal} {\bibinfo
  {journal} {Trends Ecol Evol.}\ }\textbf {\bibinfo {volume} {31}},\ \bibinfo
  {pages} {902–904} (\bibinfo {year} {2016})}\BibitemShut {NoStop}%
\bibitem [{\citenamefont {Scheffer}\ \emph
  {et~al.}(2009{\natexlab{a}})\citenamefont {Scheffer}, \citenamefont
  {Bascompte}, \citenamefont {Brock}, \citenamefont {Brovkin}, \citenamefont
  {Carpenter}, \citenamefont {Dakos}, \citenamefont {Held}, \citenamefont
  {Van~Nes}, \citenamefont {Rietkerk},\ and\ \citenamefont
  {Sugihara}}]{Schefferetal:2009}%
  \BibitemOpen
  \bibfield  {author} {\bibinfo {author} {\bibfnamefont {M.}~\bibnamefont
  {Scheffer}}, \bibinfo {author} {\bibfnamefont {J.}~\bibnamefont {Bascompte}},
  \bibinfo {author} {\bibfnamefont {W.~A.}\ \bibnamefont {Brock}}, \bibinfo
  {author} {\bibfnamefont {V.}~\bibnamefont {Brovkin}}, \bibinfo {author}
  {\bibfnamefont {S.~R.}\ \bibnamefont {Carpenter}}, \bibinfo {author}
  {\bibfnamefont {V.}~\bibnamefont {Dakos}}, \bibinfo {author} {\bibfnamefont
  {H.}~\bibnamefont {Held}}, \bibinfo {author} {\bibfnamefont {E.~H.}\
  \bibnamefont {Van~Nes}}, \bibinfo {author} {\bibfnamefont {M.}~\bibnamefont
  {Rietkerk}},\ and\ \bibinfo {author} {\bibfnamefont {G.}~\bibnamefont
  {Sugihara}},\ }\bibfield  {title} {\bibinfo {title} {Early-warning signals
  for critical transitions},\ }\href@noop {} {\bibfield  {journal} {\bibinfo
  {journal} {Nature}\ }\textbf {\bibinfo {volume} {461}},\ \bibinfo {pages}
  {53} (\bibinfo {year} {2009}{\natexlab{a}})}\BibitemShut {NoStop}%
\bibitem [{\citenamefont {Scheffer}\ \emph {et~al.}(2012)\citenamefont
  {Scheffer}, \citenamefont {Carpenter}, \citenamefont {Lenton}, \citenamefont
  {Bascompte}, \citenamefont {Brock}, \citenamefont {Dakos}, \citenamefont
  {van~de Koppel}, \citenamefont {van~de Leemput}, \citenamefont {Levin},
  \citenamefont {van Nes}, \citenamefont {Pascual},\ and\ \citenamefont
  {Vandermeer}}]{Scheffer2012}%
  \BibitemOpen
  \bibfield  {author} {\bibinfo {author} {\bibfnamefont {M.}~\bibnamefont
  {Scheffer}}, \bibinfo {author} {\bibfnamefont {S.~R.}\ \bibnamefont
  {Carpenter}}, \bibinfo {author} {\bibfnamefont {T.~M.}\ \bibnamefont
  {Lenton}}, \bibinfo {author} {\bibfnamefont {J.}~\bibnamefont {Bascompte}},
  \bibinfo {author} {\bibfnamefont {W.}~\bibnamefont {Brock}}, \bibinfo
  {author} {\bibfnamefont {V.}~\bibnamefont {Dakos}}, \bibinfo {author}
  {\bibfnamefont {J.}~\bibnamefont {van~de Koppel}}, \bibinfo {author}
  {\bibfnamefont {I.~A.}\ \bibnamefont {van~de Leemput}}, \bibinfo {author}
  {\bibfnamefont {S.~A.}\ \bibnamefont {Levin}}, \bibinfo {author}
  {\bibfnamefont {E.~H.}\ \bibnamefont {van Nes}}, \bibinfo {author}
  {\bibfnamefont {M.}~\bibnamefont {Pascual}},\ and\ \bibinfo {author}
  {\bibfnamefont {J.}~\bibnamefont {Vandermeer}},\ }\bibfield  {title}
  {\bibinfo {title} {Anticipating critical transitions},\ }\href
  {https://doi.org/10.1126/science.1225244} {\bibfield  {journal} {\bibinfo
  {journal} {Science}\ }\textbf {\bibinfo {volume} {338}},\ \bibinfo {pages}
  {344–348} (\bibinfo {year} {2012})}\BibitemShut {NoStop}%
\bibitem [{SI()}]{SI}%
  \BibitemOpen
  \href@noop {} {\bibinfo  {journal} {Supplementary Information provides
  details elaborating the results in the main text. It is helpful but not
  essential for understanding the main results of the paper. It contains the
  following: (1) background materials on previous approaches to anticipating
  critical transitions and limitations, (2) a detailed description of
  variational autoencoders, (3) a description of parameter-driven reservoir
  computing, (4) results on unsupervised learning for anticipating critical
  transition of Lorenz system with two independent bifurcation parameters, (5)
  results on anticipating critical transition with partial state observation,
  (6) results on anticipating critical transition in a chaotic ecosystem that
  violates the sparsity condition, and (7) hyperparameter Optimization}\
  }\BibitemShut {NoStop}%
\bibitem [{\citenamefont {Wang}\ \emph
  {et~al.}(2011{\natexlab{a}})\citenamefont {Wang}, \citenamefont {Yang},
  \citenamefont {Lai}, \citenamefont {Kovanis},\ and\ \citenamefont
  {Grebogi}}]{WYLKG:2011}%
  \BibitemOpen
\bibfield  {journal} {  }\bibfield  {author} {\bibinfo {author} {\bibfnamefont
  {W.-X.}\ \bibnamefont {Wang}}, \bibinfo {author} {\bibfnamefont
  {R.}~\bibnamefont {Yang}}, \bibinfo {author} {\bibfnamefont {Y.-C.}\
  \bibnamefont {Lai}}, \bibinfo {author} {\bibfnamefont {V.}~\bibnamefont
  {Kovanis}},\ and\ \bibinfo {author} {\bibfnamefont {C.}~\bibnamefont
  {Grebogi}},\ }\bibfield  {title} {\bibinfo {title} {Predicting catastrophes
  in nonlinear dynamical systems by compressive sensing},\ }\href
  {https://doi.org/10.1103/PhysRevLett.106.154101} {\bibfield  {journal}
  {\bibinfo  {journal} {Phys. Rev. Lett.}\ }\textbf {\bibinfo {volume} {106}},\
  \bibinfo {pages} {154101} (\bibinfo {year} {2011}{\natexlab{a}})}\BibitemShut
  {NoStop}%
\bibitem [{\citenamefont {Lim}\ \emph {et~al.}(2020)\citenamefont {Lim},
  \citenamefont {Theo~Giorgini}, \citenamefont {Moon},\ and\ \citenamefont
  {Wettlaufer}}]{Lim:2020}%
  \BibitemOpen
  \bibfield  {author} {\bibinfo {author} {\bibfnamefont {S.~H.}\ \bibnamefont
  {Lim}}, \bibinfo {author} {\bibfnamefont {L.}~\bibnamefont {Theo~Giorgini}},
  \bibinfo {author} {\bibfnamefont {W.}~\bibnamefont {Moon}},\ and\ \bibinfo
  {author} {\bibfnamefont {J.~S.}\ \bibnamefont {Wettlaufer}},\ }\bibfield
  {title} {\bibinfo {title} {Predicting critical transitions in multiscale
  dynamical systems using reservoir computing},\ }\href@noop {} {\bibfield
  {journal} {\bibinfo  {journal} {Chaos}\ }\textbf {\bibinfo {volume} {30}}
  (\bibinfo {year} {2020})}\BibitemShut {NoStop}%
\bibitem [{\citenamefont {Bury}\ \emph {et~al.}(2021)\citenamefont {Bury},
  \citenamefont {Sujith}, \citenamefont {Pavithran}, \citenamefont {Scheffer},
  \citenamefont {Lenton}, \citenamefont {Anand},\ and\ \citenamefont
  {Bauch}}]{BSPSLAB:2021}%
  \BibitemOpen
  \bibfield  {author} {\bibinfo {author} {\bibfnamefont {T.~M.}\ \bibnamefont
  {Bury}}, \bibinfo {author} {\bibfnamefont {R.}~\bibnamefont {Sujith}},
  \bibinfo {author} {\bibfnamefont {I.}~\bibnamefont {Pavithran}}, \bibinfo
  {author} {\bibfnamefont {M.}~\bibnamefont {Scheffer}}, \bibinfo {author}
  {\bibfnamefont {T.~M.}\ \bibnamefont {Lenton}}, \bibinfo {author}
  {\bibfnamefont {M.}~\bibnamefont {Anand}},\ and\ \bibinfo {author}
  {\bibfnamefont {C.~T.}\ \bibnamefont {Bauch}},\ }\bibfield  {title} {\bibinfo
  {title} {Deep learning for early warning signals of tipping points},\
  }\href@noop {} {\bibfield  {journal} {\bibinfo  {journal} {Proc. Natl. Acad.
  Sci. (USA)}\ }\textbf {\bibinfo {volume} {118}},\ \bibinfo {pages}
  {e2106140118} (\bibinfo {year} {2021})}\BibitemShut {NoStop}%
\bibitem [{\citenamefont {Kong}\ \emph
  {et~al.}(2021{\natexlab{a}})\citenamefont {Kong}, \citenamefont {Fan},
  \citenamefont {Grebogi},\ and\ \citenamefont {Lai}}]{KFGL:2021a}%
  \BibitemOpen
  \bibfield  {author} {\bibinfo {author} {\bibfnamefont {L.-W.}\ \bibnamefont
  {Kong}}, \bibinfo {author} {\bibfnamefont {H.-W.}\ \bibnamefont {Fan}},
  \bibinfo {author} {\bibfnamefont {C.}~\bibnamefont {Grebogi}},\ and\ \bibinfo
  {author} {\bibfnamefont {Y.-C.}\ \bibnamefont {Lai}},\ }\bibfield  {title}
  {\bibinfo {title} {Machine learning prediction of critical transition and
  system collapse},\ }\href@noop {} {\bibfield  {journal} {\bibinfo  {journal}
  {Phys. Rev. Res.}\ }\textbf {\bibinfo {volume} {3}},\ \bibinfo {pages}
  {013090} (\bibinfo {year} {2021}{\natexlab{a}})}\BibitemShut {NoStop}%
\bibitem [{\citenamefont {Kim}\ \emph {et~al.}(2021)\citenamefont {Kim},
  \citenamefont {Lu}, \citenamefont {Nozari}, \citenamefont {Pappas},\ and\
  \citenamefont {Bassett}}]{KLNPB:2021}%
  \BibitemOpen
  \bibfield  {author} {\bibinfo {author} {\bibfnamefont {J.~Z.}\ \bibnamefont
  {Kim}}, \bibinfo {author} {\bibfnamefont {Z.}~\bibnamefont {Lu}}, \bibinfo
  {author} {\bibfnamefont {E.}~\bibnamefont {Nozari}}, \bibinfo {author}
  {\bibfnamefont {G.~J.}\ \bibnamefont {Pappas}},\ and\ \bibinfo {author}
  {\bibfnamefont {D.~S.}\ \bibnamefont {Bassett}},\ }\bibfield  {title}
  {\bibinfo {title} {Teaching recurrent neural networks to infer global
  temporal structure from local examples},\ }\href@noop {} {\bibfield
  {journal} {\bibinfo  {journal} {Nat. Mach. Intell.}\ }\textbf {\bibinfo
  {volume} {3}},\ \bibinfo {pages} {316} (\bibinfo {year} {2021})}\BibitemShut
  {NoStop}%
\bibitem [{\citenamefont {Fan}\ \emph {et~al.}(2021)\citenamefont {Fan},
  \citenamefont {Kong}, \citenamefont {Lai},\ and\ \citenamefont
  {Wang}}]{FKLW:2021}%
  \BibitemOpen
  \bibfield  {author} {\bibinfo {author} {\bibfnamefont {H.}~\bibnamefont
  {Fan}}, \bibinfo {author} {\bibfnamefont {L.-W.}\ \bibnamefont {Kong}},
  \bibinfo {author} {\bibfnamefont {Y.-C.}\ \bibnamefont {Lai}},\ and\ \bibinfo
  {author} {\bibfnamefont {X.}~\bibnamefont {Wang}},\ }\bibfield  {title}
  {\bibinfo {title} {Anticipating synchronization with machine learning},\
  }\href@noop {} {\bibfield  {journal} {\bibinfo  {journal} {Phys. Rev. Res.}\
  }\textbf {\bibinfo {volume} {3}},\ \bibinfo {pages} {023237} (\bibinfo {year}
  {2021})}\BibitemShut {NoStop}%
\bibitem [{\citenamefont {Kong}\ \emph
  {et~al.}(2021{\natexlab{b}})\citenamefont {Kong}, \citenamefont {Fan},
  \citenamefont {Grebogi},\ and\ \citenamefont {Lai}}]{KFGL:2021b}%
  \BibitemOpen
  \bibfield  {author} {\bibinfo {author} {\bibfnamefont {L.-W.}\ \bibnamefont
  {Kong}}, \bibinfo {author} {\bibfnamefont {H.}~\bibnamefont {Fan}}, \bibinfo
  {author} {\bibfnamefont {C.}~\bibnamefont {Grebogi}},\ and\ \bibinfo {author}
  {\bibfnamefont {Y.-C.}\ \bibnamefont {Lai}},\ }\bibfield  {title} {\bibinfo
  {title} {Emergence of transient chaos and intermittency in machine
  learning},\ }\href@noop {} {\bibfield  {journal} {\bibinfo  {journal} {J.
  Phys. Complex.}\ }\textbf {\bibinfo {volume} {2}},\ \bibinfo {pages} {035014}
  (\bibinfo {year} {2021}{\natexlab{b}})}\BibitemShut {NoStop}%
\bibitem [{\citenamefont {Kong}\ \emph {et~al.}(2023)\citenamefont {Kong},
  \citenamefont {Weng}, \citenamefont {Glaz}, \citenamefont {Haile},\ and\
  \citenamefont {Lai}}]{KWGHL:2023}%
  \BibitemOpen
  \bibfield  {author} {\bibinfo {author} {\bibfnamefont {L.-W.}\ \bibnamefont
  {Kong}}, \bibinfo {author} {\bibfnamefont {Y.}~\bibnamefont {Weng}}, \bibinfo
  {author} {\bibfnamefont {B.}~\bibnamefont {Glaz}}, \bibinfo {author}
  {\bibfnamefont {M.}~\bibnamefont {Haile}},\ and\ \bibinfo {author}
  {\bibfnamefont {Y.-C.}\ \bibnamefont {Lai}},\ }\bibfield  {title} {\bibinfo
  {title} {Reservoir computing as digital twins for nonlinear dynamical
  systems},\ }\href@noop {} {\bibfield  {journal} {\bibinfo  {journal} {Chaos}\
  }\textbf {\bibinfo {volume} {33}},\ \bibinfo {pages} {033111} (\bibinfo
  {year} {2023})}\BibitemShut {NoStop}%
\bibitem [{\citenamefont {Jaeger}(2001)}]{Jaeger:2001}%
  \BibitemOpen
  \bibfield  {author} {\bibinfo {author} {\bibfnamefont {H.}~\bibnamefont
  {Jaeger}},\ }\bibfield  {title} {\bibinfo {title} {The ``echo state''
  approach to analysing and training recurrent neural networks-with an erratum
  note},\ }\href@noop {} {\bibfield  {journal} {\bibinfo  {journal} {GMD}\
  }\textbf {\bibinfo {volume} {148}},\ \bibinfo {pages} {13} (\bibinfo {year}
  {2001})}\BibitemShut {NoStop}%
\bibitem [{\citenamefont {Maass}\ \emph {et~al.}(2002)\citenamefont {Maass},
  \citenamefont {Natschl{\"a}ger},\ and\ \citenamefont {Markram}}]{MNM:2002}%
  \BibitemOpen
  \bibfield  {author} {\bibinfo {author} {\bibfnamefont {W.}~\bibnamefont
  {Maass}}, \bibinfo {author} {\bibfnamefont {T.}~\bibnamefont
  {Natschl{\"a}ger}},\ and\ \bibinfo {author} {\bibfnamefont {H.}~\bibnamefont
  {Markram}},\ }\bibfield  {title} {\bibinfo {title} {Real-time computing
  without stable states: A new framework for neural computation based on
  perturbations},\ }\href@noop {} {\bibfield  {journal} {\bibinfo  {journal}
  {Neural Comput.}\ }\textbf {\bibinfo {volume} {14}},\ \bibinfo {pages} {2531}
  (\bibinfo {year} {2002})}\BibitemShut {NoStop}%
\bibitem [{\citenamefont {Pathak}\ \emph {et~al.}(2018)\citenamefont {Pathak},
  \citenamefont {Hunt}, \citenamefont {Girvan}, \citenamefont {Lu},\ and\
  \citenamefont {Ott}}]{PHGLO:2018}%
  \BibitemOpen
  \bibfield  {author} {\bibinfo {author} {\bibfnamefont {J.}~\bibnamefont
  {Pathak}}, \bibinfo {author} {\bibfnamefont {B.}~\bibnamefont {Hunt}},
  \bibinfo {author} {\bibfnamefont {M.}~\bibnamefont {Girvan}}, \bibinfo
  {author} {\bibfnamefont {Z.}~\bibnamefont {Lu}},\ and\ \bibinfo {author}
  {\bibfnamefont {E.}~\bibnamefont {Ott}},\ }\bibfield  {title} {\bibinfo
  {title} {Model-free prediction of large spatiotemporally chaotic systems from
  data: {A} reservoir computing approach},\ }\href
  {https://doi.org/10.1103/PhysRevLett.120.024102} {\bibfield  {journal}
  {\bibinfo  {journal} {Phys. Rev. Lett.}\ }\textbf {\bibinfo {volume} {120}},\
  \bibinfo {pages} {024102} (\bibinfo {year} {2018})}\BibitemShut {NoStop}%
\bibitem [{\citenamefont {Bollt}(2021)}]{Bollt:2021}%
  \BibitemOpen
  \bibfield  {author} {\bibinfo {author} {\bibfnamefont {E.}~\bibnamefont
  {Bollt}},\ }\bibfield  {title} {\bibinfo {title} {On explaining the
  surprising success of reservoir computing forecaster of chaos? the universal
  machine learning dynamical system with contrast to var and dmd},\ }\href@noop
  {} {\bibfield  {journal} {\bibinfo  {journal} {Chaos}\ }\textbf {\bibinfo
  {volume} {31}},\ \bibinfo {pages} {013108} (\bibinfo {year}
  {2021})}\BibitemShut {NoStop}%
\bibitem [{\citenamefont {Gauthier}\ \emph {et~al.}(2021)\citenamefont
  {Gauthier}, \citenamefont {Bollt}, \citenamefont {Griffith},\ and\
  \citenamefont {Barbosa}}]{GBGB:2021}%
  \BibitemOpen
  \bibfield  {author} {\bibinfo {author} {\bibfnamefont {D.~J.}\ \bibnamefont
  {Gauthier}}, \bibinfo {author} {\bibfnamefont {E.}~\bibnamefont {Bollt}},
  \bibinfo {author} {\bibfnamefont {A.}~\bibnamefont {Griffith}},\ and\
  \bibinfo {author} {\bibfnamefont {W.~A.}\ \bibnamefont {Barbosa}},\
  }\bibfield  {title} {\bibinfo {title} {Next generation reservoir computing},\
  }\href@noop {} {\bibfield  {journal} {\bibinfo  {journal} {Nat. Commun.}\
  }\textbf {\bibinfo {volume} {12}},\ \bibinfo {pages} {5564} (\bibinfo {year}
  {2021})}\BibitemShut {NoStop}%
\bibitem [{\citenamefont {Zhai}\ \emph {et~al.}(2023)\citenamefont {Zhai},
  \citenamefont {Kong},\ and\ \citenamefont {Lai}}]{ZKL:2023}%
  \BibitemOpen
  \bibfield  {author} {\bibinfo {author} {\bibfnamefont {Z.-M.}\ \bibnamefont
  {Zhai}}, \bibinfo {author} {\bibfnamefont {L.-W.}\ \bibnamefont {Kong}},\
  and\ \bibinfo {author} {\bibfnamefont {Y.-C.}\ \bibnamefont {Lai}},\
  }\bibfield  {title} {\bibinfo {title} {Emergence of a resonance in machine
  learning},\ }\href {https://doi.org/10.1103/PhysRevResearch.5.033127}
  {\bibfield  {journal} {\bibinfo  {journal} {Phys. Rev. Res.}\ }\textbf
  {\bibinfo {volume} {5}},\ \bibinfo {pages} {033127} (\bibinfo {year}
  {2023})}\BibitemShut {NoStop}%
\bibitem [{\citenamefont {Kong}\ \emph {et~al.}(2024)\citenamefont {Kong},
  \citenamefont {Brewer},\ and\ \citenamefont {Lai}}]{KBL:2024}%
  \BibitemOpen
  \bibfield  {author} {\bibinfo {author} {\bibfnamefont {L.-W.}\ \bibnamefont
  {Kong}}, \bibinfo {author} {\bibfnamefont {G.~A.}\ \bibnamefont {Brewer}},\
  and\ \bibinfo {author} {\bibfnamefont {Y.-C.}\ \bibnamefont {Lai}},\
  }\bibfield  {title} {\bibinfo {title} {Reservoir-computing based associative
  memory and itinerancy for complex dynamical attractors},\ }\href@noop {}
  {\bibfield  {journal} {\bibinfo  {journal} {Nat. Commun.}\ }\textbf {\bibinfo
  {volume} {15}},\ \bibinfo {pages} {2815} (\bibinfo {year}
  {2024})}\BibitemShut {NoStop}%
\bibitem [{\citenamefont {Yan}\ \emph {et~al.}(2024)\citenamefont {Yan},
  \citenamefont {Huang}, \citenamefont {Bienstman}, \citenamefont {Tino},
  \citenamefont {Lin},\ and\ \citenamefont {Sun}}]{YHBTLS:2024}%
  \BibitemOpen
  \bibfield  {author} {\bibinfo {author} {\bibfnamefont {M.}~\bibnamefont
  {Yan}}, \bibinfo {author} {\bibfnamefont {C.}~\bibnamefont {Huang}}, \bibinfo
  {author} {\bibfnamefont {P.}~\bibnamefont {Bienstman}}, \bibinfo {author}
  {\bibfnamefont {P.}~\bibnamefont {Tino}}, \bibinfo {author} {\bibfnamefont
  {W.}~\bibnamefont {Lin}},\ and\ \bibinfo {author} {\bibfnamefont
  {J.}~\bibnamefont {Sun}},\ }\bibfield  {title} {\bibinfo {title} {Emerging
  opportunities and challenges for the future of reservoir computing},\
  }\href@noop {} {\bibfield  {journal} {\bibinfo  {journal} {Nat. Commun.}\
  }\textbf {\bibinfo {volume} {15}},\ \bibinfo {pages} {2056} (\bibinfo {year}
  {2024})}\BibitemShut {NoStop}%
\bibitem [{\citenamefont {Xiao}\ \emph {et~al.}(2021)\citenamefont {Xiao},
  \citenamefont {Kong}, \citenamefont {Sun},\ and\ \citenamefont
  {Lai}}]{XKSL:2021}%
  \BibitemOpen
  \bibfield  {author} {\bibinfo {author} {\bibfnamefont {R.}~\bibnamefont
  {Xiao}}, \bibinfo {author} {\bibfnamefont {L.-W.}\ \bibnamefont {Kong}},
  \bibinfo {author} {\bibfnamefont {Z.-K.}\ \bibnamefont {Sun}},\ and\ \bibinfo
  {author} {\bibfnamefont {Y.-C.}\ \bibnamefont {Lai}},\ }\bibfield  {title}
  {\bibinfo {title} {Predicting amplitude death with machine learning},\
  }\href@noop {} {\bibfield  {journal} {\bibinfo  {journal} {Phys. Rev. E}\
  }\textbf {\bibinfo {volume} {104}},\ \bibinfo {pages} {014205} (\bibinfo
  {year} {2021})}\BibitemShut {NoStop}%
\bibitem [{\citenamefont {Patel}\ \emph {et~al.}(2021)\citenamefont {Patel},
  \citenamefont {Canaday}, \citenamefont {Girvan}, \citenamefont {Pomerance},\
  and\ \citenamefont {Ott}}]{PCGPO:2021}%
  \BibitemOpen
  \bibfield  {author} {\bibinfo {author} {\bibfnamefont {D.}~\bibnamefont
  {Patel}}, \bibinfo {author} {\bibfnamefont {D.}~\bibnamefont {Canaday}},
  \bibinfo {author} {\bibfnamefont {M.}~\bibnamefont {Girvan}}, \bibinfo
  {author} {\bibfnamefont {A.}~\bibnamefont {Pomerance}},\ and\ \bibinfo
  {author} {\bibfnamefont {E.}~\bibnamefont {Ott}},\ }\bibfield  {title}
  {\bibinfo {title} {Using machine learning to predict statistical properties
  of non-stationary dynamical processes: System climate, regime transitions,
  and the effect of stochasticity},\ }\href@noop {} {\bibfield  {journal}
  {\bibinfo  {journal} {Chaos}\ }\textbf {\bibinfo {volume} {31}},\ \bibinfo
  {pages} {033149} (\bibinfo {year} {2021})}\BibitemShut {NoStop}%
\bibitem [{\citenamefont {Panahi}\ \emph {et~al.}(2024)\citenamefont {Panahi},
  \citenamefont {Kong}, \citenamefont {Moradi}, \citenamefont {Zhai},
  \citenamefont {Glaz}, \citenamefont {Haile},\ and\ \citenamefont
  {Lai}}]{Panahi2024}%
  \BibitemOpen
  \bibfield  {author} {\bibinfo {author} {\bibfnamefont {S.}~\bibnamefont
  {Panahi}}, \bibinfo {author} {\bibfnamefont {L.-W.}\ \bibnamefont {Kong}},
  \bibinfo {author} {\bibfnamefont {M.}~\bibnamefont {Moradi}}, \bibinfo
  {author} {\bibfnamefont {Z.-M.}\ \bibnamefont {Zhai}}, \bibinfo {author}
  {\bibfnamefont {B.}~\bibnamefont {Glaz}}, \bibinfo {author} {\bibfnamefont
  {M.}~\bibnamefont {Haile}},\ and\ \bibinfo {author} {\bibfnamefont {Y.-C.}\
  \bibnamefont {Lai}},\ }\bibfield  {title} {\bibinfo {title} {Machine-learning
  prediction of tipping},\ }\href@noop {} {\bibfield  {journal} {\bibinfo
  {journal} {arXiv preprint arXiv:2402.14877}\ } (\bibinfo {year}
  {2024})}\BibitemShut {NoStop}%
\bibitem [{\citenamefont {Lu}\ \emph {et~al.}(2020)\citenamefont {Lu},
  \citenamefont {Kim},\ and\ \citenamefont {Solja\ifmmode \check{c}\else
  \v{c}\fi{}i\ifmmode~\acute{c}\else \'{c}\fi{}}}]{Lu2020}%
  \BibitemOpen
  \bibfield  {author} {\bibinfo {author} {\bibfnamefont {P.~Y.}\ \bibnamefont
  {Lu}}, \bibinfo {author} {\bibfnamefont {S.}~\bibnamefont {Kim}},\ and\
  \bibinfo {author} {\bibfnamefont {M.}~\bibnamefont {Solja\ifmmode
  \check{c}\else \v{c}\fi{}i\ifmmode~\acute{c}\else \'{c}\fi{}}},\ }\bibfield
  {title} {\bibinfo {title} {Extracting interpretable physical parameters from
  spatiotemporal systems using unsupervised learning},\ }\href
  {https://doi.org/10.1103/PhysRevX.10.031056} {\bibfield  {journal} {\bibinfo
  {journal} {Phys. Rev. X}\ }\textbf {\bibinfo {volume} {10}},\ \bibinfo
  {pages} {031056} (\bibinfo {year} {2020})}\BibitemShut {NoStop}%
\bibitem [{\citenamefont {Lorenz}(1963)}]{lorenz1963}%
  \BibitemOpen
  \bibfield  {author} {\bibinfo {author} {\bibfnamefont {E.~N.}\ \bibnamefont
  {Lorenz}},\ }\bibfield  {title} {\bibinfo {title} {Deterministic nonperiodic
  flow},\ }\href@noop {} {\bibfield  {journal} {\bibinfo  {journal} {J. Atmos.
  Sci.}\ }\textbf {\bibinfo {volume} {20}},\ \bibinfo {pages} {130} (\bibinfo
  {year} {1963})}\BibitemShut {NoStop}%
\bibitem [{\citenamefont {Kuramoto}(1978)}]{Kuramoto:1978}%
  \BibitemOpen
  \bibfield  {author} {\bibinfo {author} {\bibfnamefont {Y.}~\bibnamefont
  {Kuramoto}},\ }\bibfield  {title} {\bibinfo {title} {Diffusion-induced chaos
  in reaction systems},\ }\href@noop {} {\bibfield  {journal} {\bibinfo
  {journal} {Prog. Theo. Phys. Supp.}\ }\textbf {\bibinfo {volume} {64}},\
  \bibinfo {pages} {346} (\bibinfo {year} {1978})}\BibitemShut {NoStop}%
\bibitem [{\citenamefont {Sivashinsky}(1980)}]{Sivashinsky:1980}%
  \BibitemOpen
  \bibfield  {author} {\bibinfo {author} {\bibfnamefont {G.~I.}\ \bibnamefont
  {Sivashinsky}},\ }\bibfield  {title} {\bibinfo {title} {On flame propagation
  under conditions of stoichiometry},\ }\href@noop {} {\bibfield  {journal}
  {\bibinfo  {journal} {SIAM J. Appl. Math.}\ }\textbf {\bibinfo {volume}
  {39}},\ \bibinfo {pages} {67} (\bibinfo {year} {1980})}\BibitemShut {NoStop}%
\bibitem [{\citenamefont {Grebogi}\ \emph {et~al.}(1983)\citenamefont
  {Grebogi}, \citenamefont {Ott},\ and\ \citenamefont {Yorke}}]{GOY:1983}%
  \BibitemOpen
  \bibfield  {author} {\bibinfo {author} {\bibfnamefont {C.}~\bibnamefont
  {Grebogi}}, \bibinfo {author} {\bibfnamefont {E.}~\bibnamefont {Ott}},\ and\
  \bibinfo {author} {\bibfnamefont {J.~A.}\ \bibnamefont {Yorke}},\ }\bibfield
  {title} {\bibinfo {title} {Crises, sudden changes in chaotic attractors, and
  transient chaos},\ }\href {https://doi.org/10.1016/0167-2789(83)90126-4}
  {\bibfield  {journal} {\bibinfo  {journal} {Physica D}\ }\textbf {\bibinfo
  {volume} {7}},\ \bibinfo {pages} {181–200} (\bibinfo {year}
  {1983})}\BibitemShut {NoStop}%
\bibitem [{\citenamefont {Crutchfield}\ and\ \citenamefont
  {McNamara}(1987)}]{CM:1987}%
  \BibitemOpen
  \bibfield  {author} {\bibinfo {author} {\bibfnamefont {J.~P.}\ \bibnamefont
  {Crutchfield}}\ and\ \bibinfo {author} {\bibfnamefont {B.}~\bibnamefont
  {McNamara}},\ }\bibfield  {title} {\bibinfo {title} {Equations of motion from
  a data series},\ }\href@noop {} {\bibfield  {journal} {\bibinfo  {journal}
  {Complex Sys.}\ }\textbf {\bibinfo {volume} {1}},\ \bibinfo {pages} {417}
  (\bibinfo {year} {1987})}\BibitemShut {NoStop}%
\bibitem [{\citenamefont {Bollt}(2000)}]{Bollt:2000}%
  \BibitemOpen
  \bibfield  {author} {\bibinfo {author} {\bibfnamefont {E.~M.}\ \bibnamefont
  {Bollt}},\ }\bibfield  {title} {\bibinfo {title} {Controlling chaos and the
  inverse frobenius-perron problem: global stabilization of arbitrary invariant
  measures},\ }\href@noop {} {\bibfield  {journal} {\bibinfo  {journal} {Int.
  J. Bif. Chaos}\ }\textbf {\bibinfo {volume} {10}},\ \bibinfo {pages} {1033}
  (\bibinfo {year} {2000})}\BibitemShut {NoStop}%
\bibitem [{\citenamefont {Yao}\ and\ \citenamefont {Bollt}(2007)}]{YB:2007}%
  \BibitemOpen
  \bibfield  {author} {\bibinfo {author} {\bibfnamefont {C.}~\bibnamefont
  {Yao}}\ and\ \bibinfo {author} {\bibfnamefont {E.~M.}\ \bibnamefont
  {Bollt}},\ }\bibfield  {title} {\bibinfo {title} {Modeling and nonlinear
  parameter estimation with {Kronecker} product representation for coupled
  oscillators and spatiotemporal systems},\ }\href@noop {} {\bibfield
  {journal} {\bibinfo  {journal} {Physica D}\ }\textbf {\bibinfo {volume}
  {227}},\ \bibinfo {pages} {78} (\bibinfo {year} {2007})}\BibitemShut
  {NoStop}%
\bibitem [{\citenamefont {R{\"o}ssler}(1976)}]{rossler1976equation}%
  \BibitemOpen
  \bibfield  {author} {\bibinfo {author} {\bibfnamefont {O.~E.}\ \bibnamefont
  {R{\"o}ssler}},\ }\bibfield  {title} {\bibinfo {title} {An equation for
  continuous chaos},\ }\href@noop {} {\bibfield  {journal} {\bibinfo  {journal}
  {Phys. Lett. A}\ }\textbf {\bibinfo {volume} {57}},\ \bibinfo {pages} {397}
  (\bibinfo {year} {1976})}\BibitemShut {NoStop}%
\bibitem [{\citenamefont {H\'{e}non}(1976)}]{Henon:1976}%
  \BibitemOpen
  \bibfield  {author} {\bibinfo {author} {\bibfnamefont {M.}~\bibnamefont
  {H\'{e}non}},\ }\bibfield  {title} {\bibinfo {title} {A two-dimensional
  mapping with a strange attractor},\ }\href@noop {} {\bibfield  {journal}
  {\bibinfo  {journal} {Commun. Math. Phys.}\ }\textbf {\bibinfo {volume}
  {50}},\ \bibinfo {pages} {69} (\bibinfo {year} {1976})}\BibitemShut {NoStop}%
\bibitem [{\citenamefont {Cand\`{e}s}\ \emph
  {et~al.}(2006{\natexlab{a}})\citenamefont {Cand\`{e}s}, \citenamefont
  {Romberg},\ and\ \citenamefont {Tao}}]{CRT:2006a}%
  \BibitemOpen
  \bibfield  {author} {\bibinfo {author} {\bibfnamefont {E.}~\bibnamefont
  {Cand\`{e}s}}, \bibinfo {author} {\bibfnamefont {J.}~\bibnamefont
  {Romberg}},\ and\ \bibinfo {author} {\bibfnamefont {T.}~\bibnamefont {Tao}},\
  }\bibfield  {title} {\bibinfo {title} {Robust uncertainty principles: exact
  signal reconstruction from highly incomplete frequency information},\
  }\href@noop {} {\bibfield  {journal} {\bibinfo  {journal} {IEEE Trans. Info.
  Theory}\ }\textbf {\bibinfo {volume} {52}},\ \bibinfo {pages} {489} (\bibinfo
  {year} {2006}{\natexlab{a}})}\BibitemShut {NoStop}%
\bibitem [{\citenamefont {Cand\`{e}s}\ \emph
  {et~al.}(2006{\natexlab{b}})\citenamefont {Cand\`{e}s}, \citenamefont
  {Romberg},\ and\ \citenamefont {Tao}}]{CRT:2006b}%
  \BibitemOpen
  \bibfield  {author} {\bibinfo {author} {\bibfnamefont {E.}~\bibnamefont
  {Cand\`{e}s}}, \bibinfo {author} {\bibfnamefont {J.}~\bibnamefont
  {Romberg}},\ and\ \bibinfo {author} {\bibfnamefont {T.}~\bibnamefont {Tao}},\
  }\bibfield  {title} {\bibinfo {title} {Stable signal recovery from incomplete
  and inaccurate measurements},\ }\href@noop {} {\bibfield  {journal} {\bibinfo
   {journal} {Comm. Pure Appl. Math.}\ }\textbf {\bibinfo {volume} {59}},\
  \bibinfo {pages} {1207} (\bibinfo {year} {2006}{\natexlab{b}})}\BibitemShut
  {NoStop}%
\bibitem [{\citenamefont {Donoho}(2006)}]{Donoho:2006}%
  \BibitemOpen
  \bibfield  {author} {\bibinfo {author} {\bibfnamefont {D.}~\bibnamefont
  {Donoho}},\ }\bibfield  {title} {\bibinfo {title} {Compressed sensing},\
  }\href@noop {} {\bibfield  {journal} {\bibinfo  {journal} {IEEE Trans. Info.
  Theory}\ }\textbf {\bibinfo {volume} {52}},\ \bibinfo {pages} {1289}
  (\bibinfo {year} {2006})}\BibitemShut {NoStop}%
\bibitem [{\citenamefont {Baraniuk}(2007)}]{Baraniuk:2007}%
  \BibitemOpen
  \bibfield  {author} {\bibinfo {author} {\bibfnamefont {R.~G.}\ \bibnamefont
  {Baraniuk}},\ }\bibfield  {title} {\bibinfo {title} {Compressed sensing},\
  }\href@noop {} {\bibfield  {journal} {\bibinfo  {journal} {IEEE Signal
  Process. Mag.}\ }\textbf {\bibinfo {volume} {24}},\ \bibinfo {pages} {118}
  (\bibinfo {year} {2007})}\BibitemShut {NoStop}%
\bibitem [{\citenamefont {Cande\`s}\ and\ \citenamefont
  {Wakin}(2008)}]{CW:2008}%
  \BibitemOpen
  \bibfield  {author} {\bibinfo {author} {\bibfnamefont {E.}~\bibnamefont
  {Cande\`s}}\ and\ \bibinfo {author} {\bibfnamefont {M.}~\bibnamefont
  {Wakin}},\ }\bibfield  {title} {\bibinfo {title} {An introduction to
  compressive sampling},\ }\href@noop {} {\bibfield  {journal} {\bibinfo
  {journal} {IEEE Signal Process. Mag.}\ }\textbf {\bibinfo {volume} {25}},\
  \bibinfo {pages} {21} (\bibinfo {year} {2008})}\BibitemShut {NoStop}%
\bibitem [{\citenamefont {Wang}\ \emph {et~al.}(2016)\citenamefont {Wang},
  \citenamefont {Lai},\ and\ \citenamefont {Grebogi}}]{WLG:2016}%
  \BibitemOpen
  \bibfield  {author} {\bibinfo {author} {\bibfnamefont {W.}~\bibnamefont
  {Wang}}, \bibinfo {author} {\bibfnamefont {Y.-C.}\ \bibnamefont {Lai}},\ and\
  \bibinfo {author} {\bibfnamefont {C.}~\bibnamefont {Grebogi}},\ }\bibfield
  {title} {\bibinfo {title} {Data based identification and prediction of
  nonlinear and complex dynamical systems},\ }\href@noop {} {\bibfield
  {journal} {\bibinfo  {journal} {Phys. Rep.}\ }\textbf {\bibinfo {volume}
  {644}},\ \bibinfo {pages} {1} (\bibinfo {year} {2016})}\BibitemShut {NoStop}%
\bibitem [{\citenamefont {Lai}(2021)}]{Lai:2021}%
  \BibitemOpen
  \bibfield  {author} {\bibinfo {author} {\bibfnamefont {Y.-C.}\ \bibnamefont
  {Lai}},\ }\bibfield  {title} {\bibinfo {title} {Finding nonlinear system
  equations and complex network structures from data: A sparse optimization
  approach},\ }\href@noop {} {\bibfield  {journal} {\bibinfo  {journal}
  {Chaos}\ }\textbf {\bibinfo {volume} {31}},\ \bibinfo {pages} {082101}
  (\bibinfo {year} {2021})}\BibitemShut {NoStop}%
\bibitem [{\citenamefont {Wang}\ \emph
  {et~al.}(2011{\natexlab{b}})\citenamefont {Wang}, \citenamefont {Yang},
  \citenamefont {Lai}, \citenamefont {Kovanis},\ and\ \citenamefont
  {Harrison}}]{WYLKH:2011}%
  \BibitemOpen
  \bibfield  {author} {\bibinfo {author} {\bibfnamefont {W.-X.}\ \bibnamefont
  {Wang}}, \bibinfo {author} {\bibfnamefont {R.}~\bibnamefont {Yang}}, \bibinfo
  {author} {\bibfnamefont {Y.-C.}\ \bibnamefont {Lai}}, \bibinfo {author}
  {\bibfnamefont {V.}~\bibnamefont {Kovanis}},\ and\ \bibinfo {author}
  {\bibfnamefont {M.~A.~F.}\ \bibnamefont {Harrison}},\ }\bibfield  {title}
  {\bibinfo {title} {Time-series-based prediction of complex oscillator
  networks via compressive sensing},\ }\href@noop {} {\bibfield  {journal}
  {\bibinfo  {journal} {EPL (Europhys. Lett.)}\ }\textbf {\bibinfo {volume}
  {94}},\ \bibinfo {pages} {48006} (\bibinfo {year}
  {2011}{\natexlab{b}})}\BibitemShut {NoStop}%
\bibitem [{\citenamefont {Wang}\ \emph
  {et~al.}(2011{\natexlab{c}})\citenamefont {Wang}, \citenamefont {Lai},
  \citenamefont {Grebogi},\ and\ \citenamefont {Ye}}]{WLGY:2011}%
  \BibitemOpen
  \bibfield  {author} {\bibinfo {author} {\bibfnamefont {W.-X.}\ \bibnamefont
  {Wang}}, \bibinfo {author} {\bibfnamefont {Y.-C.}\ \bibnamefont {Lai}},
  \bibinfo {author} {\bibfnamefont {C.}~\bibnamefont {Grebogi}},\ and\ \bibinfo
  {author} {\bibfnamefont {J.-P.}\ \bibnamefont {Ye}},\ }\bibfield  {title}
  {\bibinfo {title} {Network reconstruction based on evolutionary-game data via
  compressive sensing},\ }\href@noop {} {\bibfield  {journal} {\bibinfo
  {journal} {Phys. Rev. X}\ }\textbf {\bibinfo {volume} {1}},\ \bibinfo {pages}
  {021021} (\bibinfo {year} {2011}{\natexlab{c}})}\BibitemShut {NoStop}%
\bibitem [{\citenamefont {Su}\ \emph {et~al.}(2012{\natexlab{a}})\citenamefont
  {Su}, \citenamefont {Ni}, \citenamefont {Wang},\ and\ \citenamefont
  {Lai}}]{SNWL:2012}%
  \BibitemOpen
  \bibfield  {author} {\bibinfo {author} {\bibfnamefont {R.-Q.}\ \bibnamefont
  {Su}}, \bibinfo {author} {\bibfnamefont {X.}~\bibnamefont {Ni}}, \bibinfo
  {author} {\bibfnamefont {W.-X.}\ \bibnamefont {Wang}},\ and\ \bibinfo
  {author} {\bibfnamefont {Y.-C.}\ \bibnamefont {Lai}},\ }\bibfield  {title}
  {\bibinfo {title} {Forecasting synchronizability of complex networks from
  data},\ }\href@noop {} {\bibfield  {journal} {\bibinfo  {journal} {Phys. Rev.
  E}\ }\textbf {\bibinfo {volume} {85}},\ \bibinfo {pages} {056220} (\bibinfo
  {year} {2012}{\natexlab{a}})}\BibitemShut {NoStop}%
\bibitem [{\citenamefont {Su}\ \emph {et~al.}(2012{\natexlab{b}})\citenamefont
  {Su}, \citenamefont {Wang},\ and\ \citenamefont {Lai}}]{SWL:2012}%
  \BibitemOpen
  \bibfield  {author} {\bibinfo {author} {\bibfnamefont {R.-Q.}\ \bibnamefont
  {Su}}, \bibinfo {author} {\bibfnamefont {W.-X.}\ \bibnamefont {Wang}},\ and\
  \bibinfo {author} {\bibfnamefont {Y.-C.}\ \bibnamefont {Lai}},\ }\bibfield
  {title} {\bibinfo {title} {Detecting hidden nodes in complex networks from
  time series},\ }\href@noop {} {\bibfield  {journal} {\bibinfo  {journal}
  {Phys. Rev. E}\ }\textbf {\bibinfo {volume} {85}},\ \bibinfo {pages} {065201}
  (\bibinfo {year} {2012}{\natexlab{b}})}\BibitemShut {NoStop}%
\bibitem [{\citenamefont {Su}\ \emph {et~al.}(2014)\citenamefont {Su},
  \citenamefont {Lai}, \citenamefont {Wang},\ and\ \citenamefont
  {Do}}]{SLWD:2014}%
  \BibitemOpen
  \bibfield  {author} {\bibinfo {author} {\bibfnamefont {R.-Q.}\ \bibnamefont
  {Su}}, \bibinfo {author} {\bibfnamefont {Y.-C.}\ \bibnamefont {Lai}},
  \bibinfo {author} {\bibfnamefont {X.}~\bibnamefont {Wang}},\ and\ \bibinfo
  {author} {\bibfnamefont {Y.-H.}\ \bibnamefont {Do}},\ }\bibfield  {title}
  {\bibinfo {title} {Uncovering hidden nodes in complex networks in the
  presence of noise},\ }\href@noop {} {\bibfield  {journal} {\bibinfo
  {journal} {Sci. Rep.}\ }\textbf {\bibinfo {volume} {4}},\ \bibinfo {pages}
  {3944} (\bibinfo {year} {2014})}\BibitemShut {NoStop}%
\bibitem [{\citenamefont {Shen}\ \emph {et~al.}(2014)\citenamefont {Shen},
  \citenamefont {Wang}, \citenamefont {Fan}, \citenamefont {Di},\ and\
  \citenamefont {Lai}}]{SWFDL:2014}%
  \BibitemOpen
  \bibfield  {author} {\bibinfo {author} {\bibfnamefont {Z.}~\bibnamefont
  {Shen}}, \bibinfo {author} {\bibfnamefont {W.-X.}\ \bibnamefont {Wang}},
  \bibinfo {author} {\bibfnamefont {Y.}~\bibnamefont {Fan}}, \bibinfo {author}
  {\bibfnamefont {Z.}~\bibnamefont {Di}},\ and\ \bibinfo {author}
  {\bibfnamefont {Y.-C.}\ \bibnamefont {Lai}},\ }\bibfield  {title} {\bibinfo
  {title} {Reconstructing propagation networks with natural diversity and
  identifying hidden sources},\ }\href@noop {} {\bibfield  {journal} {\bibinfo
  {journal} {Nat. Commun.}\ }\textbf {\bibinfo {volume} {5}},\ \bibinfo {pages}
  {4323} (\bibinfo {year} {2014})}\BibitemShut {NoStop}%
\bibitem [{\citenamefont {Scheffer}(2004)}]{Scheffer:2004}%
  \BibitemOpen
  \bibfield  {author} {\bibinfo {author} {\bibfnamefont {M.}~\bibnamefont
  {Scheffer}},\ }\href@noop {} {\emph {\bibinfo {title} {{Ecology of Shallow
  Lakes}}}}\ (\bibinfo  {publisher} {Springer Science \& Business Media},\
  \bibinfo {year} {2004})\BibitemShut {NoStop}%
\bibitem [{\citenamefont {Scheffer}(2010)}]{Scheffer:2010}%
  \BibitemOpen
  \bibfield  {author} {\bibinfo {author} {\bibfnamefont {M.}~\bibnamefont
  {Scheffer}},\ }\bibfield  {title} {\bibinfo {title} {Complex systems:
  foreseeing tipping points},\ }\href@noop {} {\bibfield  {journal} {\bibinfo
  {journal} {Nature}\ }\textbf {\bibinfo {volume} {467}},\ \bibinfo {pages}
  {411} (\bibinfo {year} {2010})}\BibitemShut {NoStop}%
\bibitem [{\citenamefont {Wysham}\ and\ \citenamefont
  {Hastings}(2010)}]{WH:2010}%
  \BibitemOpen
  \bibfield  {author} {\bibinfo {author} {\bibfnamefont {D.~B.}\ \bibnamefont
  {Wysham}}\ and\ \bibinfo {author} {\bibfnamefont {A.}~\bibnamefont
  {Hastings}},\ }\bibfield  {title} {\bibinfo {title} {Regime shifts in
  ecological systems can occur with no warning},\ }\href@noop {} {\bibfield
  {journal} {\bibinfo  {journal} {Ecol. Lett.}\ }\textbf {\bibinfo {volume}
  {13}},\ \bibinfo {pages} {464} (\bibinfo {year} {2010})}\BibitemShut
  {NoStop}%
\bibitem [{\citenamefont {Tylianakis}\ and\ \citenamefont
  {Coux}(2014)}]{TC:2014}%
  \BibitemOpen
  \bibfield  {author} {\bibinfo {author} {\bibfnamefont {J.~M.}\ \bibnamefont
  {Tylianakis}}\ and\ \bibinfo {author} {\bibfnamefont {C.}~\bibnamefont
  {Coux}},\ }\bibfield  {title} {\bibinfo {title} {Tipping points in ecological
  networks},\ }\href@noop {} {\bibfield  {journal} {\bibinfo  {journal}
  {Trends. Plant. Sci.}\ }\textbf {\bibinfo {volume} {19}},\ \bibinfo {pages}
  {281} (\bibinfo {year} {2014})}\BibitemShut {NoStop}%
\bibitem [{\citenamefont {Jiang}\ \emph {et~al.}(2018)\citenamefont {Jiang},
  \citenamefont {Huang}, \citenamefont {Seager}, \citenamefont {Lin},
  \citenamefont {Grebogi}, \citenamefont {Hastings},\ and\ \citenamefont
  {Lai}}]{JHSLGHL:2018}%
  \BibitemOpen
  \bibfield  {author} {\bibinfo {author} {\bibfnamefont {J.}~\bibnamefont
  {Jiang}}, \bibinfo {author} {\bibfnamefont {Z.-G.}\ \bibnamefont {Huang}},
  \bibinfo {author} {\bibfnamefont {T.~P.}\ \bibnamefont {Seager}}, \bibinfo
  {author} {\bibfnamefont {W.}~\bibnamefont {Lin}}, \bibinfo {author}
  {\bibfnamefont {C.}~\bibnamefont {Grebogi}}, \bibinfo {author} {\bibfnamefont
  {A.}~\bibnamefont {Hastings}},\ and\ \bibinfo {author} {\bibfnamefont
  {Y.-C.}\ \bibnamefont {Lai}},\ }\bibfield  {title} {\bibinfo {title}
  {Predicting tipping points in mutualistic networks through dimension
  reduction},\ }\href@noop {} {\bibfield  {journal} {\bibinfo  {journal} {Proc.
  Nat. Acad. Sci. (USA)}\ }\textbf {\bibinfo {volume} {115}},\ \bibinfo {pages}
  {E639} (\bibinfo {year} {2018})}\BibitemShut {NoStop}%
\bibitem [{\citenamefont {Yang}\ \emph {et~al.}(2018)\citenamefont {Yang},
  \citenamefont {Li}, \citenamefont {Tang}, \citenamefont {Liu}, \citenamefont
  {Zhang}, \citenamefont {Chen},\ and\ \citenamefont {Xia}}]{YLTLZCX:2018}%
  \BibitemOpen
  \bibfield  {author} {\bibinfo {author} {\bibfnamefont {B.}~\bibnamefont
  {Yang}}, \bibinfo {author} {\bibfnamefont {M.}~\bibnamefont {Li}}, \bibinfo
  {author} {\bibfnamefont {W.}~\bibnamefont {Tang}}, \bibinfo {author}
  {\bibfnamefont {W.}~\bibnamefont {Liu}}, \bibinfo {author} {\bibfnamefont
  {S.}~\bibnamefont {Zhang}}, \bibinfo {author} {\bibfnamefont
  {L.}~\bibnamefont {Chen}},\ and\ \bibinfo {author} {\bibfnamefont
  {J.}~\bibnamefont {Xia}},\ }\bibfield  {title} {\bibinfo {title} {Dynamic
  network biomarker indicates pulmonary metastasis at the tipping point of
  hepatocellular carcinoma},\ }\href@noop {} {\bibfield  {journal} {\bibinfo
  {journal} {Nat. Commun.}\ }\textbf {\bibinfo {volume} {9}},\ \bibinfo {pages}
  {678} (\bibinfo {year} {2018})}\BibitemShut {NoStop}%
\bibitem [{\citenamefont {Jiang}\ \emph {et~al.}(2019)\citenamefont {Jiang},
  \citenamefont {Hastings},\ and\ \citenamefont {Lai}}]{JHL:2019}%
  \BibitemOpen
  \bibfield  {author} {\bibinfo {author} {\bibfnamefont {J.}~\bibnamefont
  {Jiang}}, \bibinfo {author} {\bibfnamefont {A.}~\bibnamefont {Hastings}},\
  and\ \bibinfo {author} {\bibfnamefont {Y.-C.}\ \bibnamefont {Lai}},\
  }\bibfield  {title} {\bibinfo {title} {Harnessing tipping points in complex
  ecological networks},\ }\href@noop {} {\bibfield  {journal} {\bibinfo
  {journal} {J. R. Soc. Interface}\ }\textbf {\bibinfo {volume} {16}},\
  \bibinfo {pages} {20190345} (\bibinfo {year} {2019})}\BibitemShut {NoStop}%
\bibitem [{\citenamefont {Meng}\ \emph {et~al.}(2020)\citenamefont {Meng},
  \citenamefont {Lai},\ and\ \citenamefont {Grebogi}}]{meng2020tipping}%
  \BibitemOpen
  \bibfield  {author} {\bibinfo {author} {\bibfnamefont {Y.}~\bibnamefont
  {Meng}}, \bibinfo {author} {\bibfnamefont {Y.-C.}\ \bibnamefont {Lai}},\ and\
  \bibinfo {author} {\bibfnamefont {C.}~\bibnamefont {Grebogi}},\ }\bibfield
  {title} {\bibinfo {title} {Tipping point and noise-induced transients in
  ecological networks},\ }\href@noop {} {\bibfield  {journal} {\bibinfo
  {journal} {J. R. Soc. Interface.}\ }\textbf {\bibinfo {volume} {17}},\
  \bibinfo {pages} {20200645} (\bibinfo {year} {2020})}\BibitemShut {NoStop}%
\bibitem [{\citenamefont {Meng}\ \emph {et~al.}(2022)\citenamefont {Meng},
  \citenamefont {Lai},\ and\ \citenamefont {Grebogi}}]{MLG:2022}%
  \BibitemOpen
  \bibfield  {author} {\bibinfo {author} {\bibfnamefont {Y.}~\bibnamefont
  {Meng}}, \bibinfo {author} {\bibfnamefont {Y.-C.}\ \bibnamefont {Lai}},\ and\
  \bibinfo {author} {\bibfnamefont {C.}~\bibnamefont {Grebogi}},\ }\bibfield
  {title} {\bibinfo {title} {The fundamental benefits of multiplexity in
  ecological networks},\ }\href@noop {} {\bibfield  {journal} {\bibinfo
  {journal} {J. R. Soc. Interface}\ }\textbf {\bibinfo {volume} {19}},\
  \bibinfo {pages} {20220438} (\bibinfo {year} {2022})}\BibitemShut {NoStop}%
\bibitem [{\citenamefont {Scheffer}\ \emph
  {et~al.}(2009{\natexlab{b}})\citenamefont {Scheffer}, \citenamefont
  {Bascompte}, \citenamefont {Brock}, \citenamefont {Brovkin}, \citenamefont
  {Carpenter}, \citenamefont {Dakos}, \citenamefont {Held}, \citenamefont
  {Van~Nes}, \citenamefont {Rietkerk},\ and\ \citenamefont
  {Sugihara}}]{SBBBCDHNRS:2009}%
  \BibitemOpen
  \bibfield  {author} {\bibinfo {author} {\bibfnamefont {M.}~\bibnamefont
  {Scheffer}}, \bibinfo {author} {\bibfnamefont {J.}~\bibnamefont {Bascompte}},
  \bibinfo {author} {\bibfnamefont {W.~A.}\ \bibnamefont {Brock}}, \bibinfo
  {author} {\bibfnamefont {V.}~\bibnamefont {Brovkin}}, \bibinfo {author}
  {\bibfnamefont {S.~R.}\ \bibnamefont {Carpenter}}, \bibinfo {author}
  {\bibfnamefont {V.}~\bibnamefont {Dakos}}, \bibinfo {author} {\bibfnamefont
  {H.}~\bibnamefont {Held}}, \bibinfo {author} {\bibfnamefont {E.~H.}\
  \bibnamefont {Van~Nes}}, \bibinfo {author} {\bibfnamefont {M.}~\bibnamefont
  {Rietkerk}},\ and\ \bibinfo {author} {\bibfnamefont {G.}~\bibnamefont
  {Sugihara}},\ }\bibfield  {title} {\bibinfo {title} {Early-warning signals
  for critical transitions},\ }\href@noop {} {\bibfield  {journal} {\bibinfo
  {journal} {Nature}\ }\textbf {\bibinfo {volume} {461}},\ \bibinfo {pages}
  {53} (\bibinfo {year} {2009}{\natexlab{b}})}\BibitemShut {NoStop}%
\bibitem [{\citenamefont {Drake}\ and\ \citenamefont
  {Griffen}(2010)}]{DG:2010}%
  \BibitemOpen
  \bibfield  {author} {\bibinfo {author} {\bibfnamefont {J.~M.}\ \bibnamefont
  {Drake}}\ and\ \bibinfo {author} {\bibfnamefont {B.~D.}\ \bibnamefont
  {Griffen}},\ }\bibfield  {title} {\bibinfo {title} {Early warning signals of
  extinction in deteriorating environments},\ }\href@noop {} {\bibfield
  {journal} {\bibinfo  {journal} {Nature}\ }\textbf {\bibinfo {volume} {467}},\
  \bibinfo {pages} {456} (\bibinfo {year} {2010})}\BibitemShut {NoStop}%
\bibitem [{\citenamefont {Boettiger}\ and\ \citenamefont
  {Hastings}(2012)}]{BH:2012}%
  \BibitemOpen
  \bibfield  {author} {\bibinfo {author} {\bibfnamefont {C.}~\bibnamefont
  {Boettiger}}\ and\ \bibinfo {author} {\bibfnamefont {A.}~\bibnamefont
  {Hastings}},\ }\bibfield  {title} {\bibinfo {title} {Quantifying limits to
  detection of early warning for critical transitions},\ }\href@noop {}
  {\bibfield  {journal} {\bibinfo  {journal} {J. R. Soc. Interface}\ }\textbf
  {\bibinfo {volume} {9}},\ \bibinfo {pages} {2527} (\bibinfo {year}
  {2012})}\BibitemShut {NoStop}%
\bibitem [{\citenamefont {Chen}\ \emph {et~al.}(2012)\citenamefont {Chen},
  \citenamefont {Liu}, \citenamefont {Liu}, \citenamefont {Li},\ and\
  \citenamefont {Aihara}}]{CLLLA:2012}%
  \BibitemOpen
  \bibfield  {author} {\bibinfo {author} {\bibfnamefont {L.}~\bibnamefont
  {Chen}}, \bibinfo {author} {\bibfnamefont {R.}~\bibnamefont {Liu}}, \bibinfo
  {author} {\bibfnamefont {Z.-P.}\ \bibnamefont {Liu}}, \bibinfo {author}
  {\bibfnamefont {M.}~\bibnamefont {Li}},\ and\ \bibinfo {author}
  {\bibfnamefont {K.}~\bibnamefont {Aihara}},\ }\bibfield  {title} {\bibinfo
  {title} {Detecting early-warning signals for sudden deterioration of complex
  diseases by dynamical network biomarkers},\ }\href@noop {} {\bibfield
  {journal} {\bibinfo  {journal} {Sci. Rep.}\ }\textbf {\bibinfo {volume}
  {2}},\ \bibinfo {pages} {342} (\bibinfo {year} {2012})}\BibitemShut {NoStop}%
\bibitem [{\citenamefont {Dai}\ \emph {et~al.}(2012)\citenamefont {Dai},
  \citenamefont {Vorselen}, \citenamefont {Korolev},\ and\ \citenamefont
  {Gore}}]{DVKG:2012}%
  \BibitemOpen
  \bibfield  {author} {\bibinfo {author} {\bibfnamefont {L.}~\bibnamefont
  {Dai}}, \bibinfo {author} {\bibfnamefont {D.}~\bibnamefont {Vorselen}},
  \bibinfo {author} {\bibfnamefont {K.~S.}\ \bibnamefont {Korolev}},\ and\
  \bibinfo {author} {\bibfnamefont {J.}~\bibnamefont {Gore}},\ }\bibfield
  {title} {\bibinfo {title} {Generic indicators for loss of resilience before a
  tipping point leading to population collapse},\ }\href@noop {} {\bibfield
  {journal} {\bibinfo  {journal} {Science}\ }\textbf {\bibinfo {volume}
  {336}},\ \bibinfo {pages} {1175} (\bibinfo {year} {2012})}\BibitemShut
  {NoStop}%
\bibitem [{\citenamefont {Boettiger}\ \emph {et~al.}(2013)\citenamefont
  {Boettiger}, \citenamefont {Ross},\ and\ \citenamefont
  {Hastings}}]{BRH:2013}%
  \BibitemOpen
  \bibfield  {author} {\bibinfo {author} {\bibfnamefont {C.}~\bibnamefont
  {Boettiger}}, \bibinfo {author} {\bibfnamefont {N.}~\bibnamefont {Ross}},\
  and\ \bibinfo {author} {\bibfnamefont {A.}~\bibnamefont {Hastings}},\
  }\bibfield  {title} {\bibinfo {title} {Early warning signals: the charted and
  uncharted territories},\ }\href@noop {} {\bibfield  {journal} {\bibinfo
  {journal} {Theor. Ecol.}\ }\textbf {\bibinfo {volume} {6}},\ \bibinfo {pages}
  {255} (\bibinfo {year} {2013})}\BibitemShut {NoStop}%
\bibitem [{\citenamefont {van~de Leemput}\ \emph {et~al.}(2014)\citenamefont
  {van~de Leemput}, \citenamefont {Wichers}, \citenamefont {Cramer},
  \citenamefont {Borsboom}, \citenamefont {Tuerlinckx}, \citenamefont
  {Kuppens}, \citenamefont {van Nes}, \citenamefont {Viechtbauer},
  \citenamefont {Giltay}, \citenamefont {Aggen} \emph
  {et~al.}}]{LeemputEtal:2014}%
  \BibitemOpen
  \bibfield  {author} {\bibinfo {author} {\bibfnamefont {I.~A.}\ \bibnamefont
  {van~de Leemput}}, \bibinfo {author} {\bibfnamefont {M.}~\bibnamefont
  {Wichers}}, \bibinfo {author} {\bibfnamefont {A.~O.}\ \bibnamefont {Cramer}},
  \bibinfo {author} {\bibfnamefont {D.}~\bibnamefont {Borsboom}}, \bibinfo
  {author} {\bibfnamefont {F.}~\bibnamefont {Tuerlinckx}}, \bibinfo {author}
  {\bibfnamefont {P.}~\bibnamefont {Kuppens}}, \bibinfo {author} {\bibfnamefont
  {E.~H.}\ \bibnamefont {van Nes}}, \bibinfo {author} {\bibfnamefont
  {W.}~\bibnamefont {Viechtbauer}}, \bibinfo {author} {\bibfnamefont {E.~J.}\
  \bibnamefont {Giltay}}, \bibinfo {author} {\bibfnamefont {S.~H.}\
  \bibnamefont {Aggen}}, \emph {et~al.},\ }\bibfield  {title} {\bibinfo {title}
  {Critical slowing down as early warning for the onset and termination of
  depression},\ }\href@noop {} {\bibfield  {journal} {\bibinfo  {journal}
  {Proc. Natl. Acad. Sci. (USA)}\ }\textbf {\bibinfo {volume} {111}},\ \bibinfo
  {pages} {87} (\bibinfo {year} {2014})}\BibitemShut {NoStop}%
\bibitem [{\citenamefont {Boers}(2018)}]{Boers:2018}%
  \BibitemOpen
  \bibfield  {author} {\bibinfo {author} {\bibfnamefont {N.}~\bibnamefont
  {Boers}},\ }\bibfield  {title} {\bibinfo {title} {Early-warning signals for
  dansgaard-oeschger events in a high-resolution ice core record},\ }\href@noop
  {} {\bibfield  {journal} {\bibinfo  {journal} {Nat. Commun.}\ }\textbf
  {\bibinfo {volume} {9}},\ \bibinfo {pages} {1} (\bibinfo {year}
  {2018})}\BibitemShut {NoStop}%
\bibitem [{\citenamefont {Zhang}\ \emph {et~al.}(2021)\citenamefont {Zhang},
  \citenamefont {Tino}, \citenamefont {Leonardis},\ and\ \citenamefont
  {Tang}}]{Zhang2021}%
  \BibitemOpen
  \bibfield  {author} {\bibinfo {author} {\bibfnamefont {Y.}~\bibnamefont
  {Zhang}}, \bibinfo {author} {\bibfnamefont {P.}~\bibnamefont {Tino}},
  \bibinfo {author} {\bibfnamefont {A.}~\bibnamefont {Leonardis}},\ and\
  \bibinfo {author} {\bibfnamefont {K.}~\bibnamefont {Tang}},\ }\bibfield
  {title} {\bibinfo {title} {A survey on neural network interpretability},\
  }\href {https://doi.org/10.1109/tetci.2021.3100641} {\bibfield  {journal}
  {\bibinfo  {journal} {IEEE Trans. Emerg. Top. Comput. Intell.}\ }\textbf
  {\bibinfo {volume} {5}},\ \bibinfo {pages} {726–742} (\bibinfo {year}
  {2021})}\BibitemShut {NoStop}%
\bibitem [{\citenamefont {Rudin}(2019)}]{Rudin2019}%
  \BibitemOpen
  \bibfield  {author} {\bibinfo {author} {\bibfnamefont {C.}~\bibnamefont
  {Rudin}},\ }\bibfield  {title} {\bibinfo {title} {Stop explaining black box
  machine learning models for high stakes decisions and use interpretable
  models instead},\ }\href {https://doi.org/10.1038/s42256-019-0048-x}
  {\bibfield  {journal} {\bibinfo  {journal} {Nat. Mach. Intell.}\ }\textbf
  {\bibinfo {volume} {1}},\ \bibinfo {pages} {206–215} (\bibinfo {year}
  {2019})}\BibitemShut {NoStop}%
\bibitem [{\citenamefont {Alao}\ \emph {et~al.}(2021)\citenamefont {Alao},
  \citenamefont {Lu},\ and\ \citenamefont {Soljacic}}]{Alao2021}%
  \BibitemOpen
  \bibfield  {author} {\bibinfo {author} {\bibfnamefont {O.}~\bibnamefont
  {Alao}}, \bibinfo {author} {\bibfnamefont {P.~Y.}\ \bibnamefont {Lu}},\ and\
  \bibinfo {author} {\bibfnamefont {M.}~\bibnamefont {Soljacic}},\ }\bibfield
  {title} {\bibinfo {title} {Discovering dynamical parameters by interpreting
  echo state networks},\ }in\ \href@noop {} {\emph {\bibinfo {booktitle}
  {NeurIPS 2021 AI for Science Workshop}}}\ (\bibinfo {year}
  {2021})\BibitemShut {NoStop}%
\bibitem [{\citenamefont {Goodfellow}\ \emph {et~al.}(2016)\citenamefont
  {Goodfellow}, \citenamefont {Bengio},\ and\ \citenamefont
  {Courville}}]{Goodfello2016}%
  \BibitemOpen
  \bibfield  {author} {\bibinfo {author} {\bibfnamefont {I.}~\bibnamefont
  {Goodfellow}}, \bibinfo {author} {\bibfnamefont {Y.}~\bibnamefont {Bengio}},\
  and\ \bibinfo {author} {\bibfnamefont {A.}~\bibnamefont {Courville}},\ }\href
  {https://books.google.com/books?id=omivDQAAQBAJ} {\emph {\bibinfo {title}
  {Deep Learning}}},\ Adaptive Computation and Machine Learning series\
  (\bibinfo  {publisher} {MIT Press},\ \bibinfo {year} {2016})\BibitemShut
  {NoStop}%
\bibitem [{\citenamefont {Kingma}\ and\ \citenamefont
  {Welling}(2013)}]{Kingma2013}%
  \BibitemOpen
  \bibfield  {author} {\bibinfo {author} {\bibfnamefont {D.~P.}\ \bibnamefont
  {Kingma}}\ and\ \bibinfo {author} {\bibfnamefont {M.}~\bibnamefont
  {Welling}},\ }\bibfield  {title} {\bibinfo {title} {Auto-encoding variational
  bayes},\ }\href@noop {} {\bibfield  {journal} {\bibinfo  {journal} {arXiv
  preprint arXiv:1312.6114}\ } (\bibinfo {year} {2013})}\BibitemShut {NoStop}%
\bibitem [{\citenamefont {Rezende}\ \emph {et~al.}(2014)\citenamefont
  {Rezende}, \citenamefont {Mohamed},\ and\ \citenamefont
  {Wierstra}}]{rezende2014stochastic}%
  \BibitemOpen
  \bibfield  {author} {\bibinfo {author} {\bibfnamefont {D.~J.}\ \bibnamefont
  {Rezende}}, \bibinfo {author} {\bibfnamefont {S.}~\bibnamefont {Mohamed}},\
  and\ \bibinfo {author} {\bibfnamefont {D.}~\bibnamefont {Wierstra}},\
  }\bibfield  {title} {\bibinfo {title} {Stochastic backpropagation and
  approximate inference in deep generative models},\ }in\ \href@noop {} {\emph
  {\bibinfo {booktitle} {International Conference on Machine Learning}}}\
  (\bibinfo {organization} {PMLR},\ \bibinfo {year} {2014})\ pp.\ \bibinfo
  {pages} {1278--1286}\BibitemShut {NoStop}%
\bibitem [{\citenamefont {Manjunath}\ and\ \citenamefont
  {Jaeger}(2013)}]{MJ:2013}%
  \BibitemOpen
  \bibfield  {author} {\bibinfo {author} {\bibfnamefont {G.}~\bibnamefont
  {Manjunath}}\ and\ \bibinfo {author} {\bibfnamefont {H.}~\bibnamefont
  {Jaeger}},\ }\bibfield  {title} {\bibinfo {title} {Echo state property linked
  to an input: Exploring a fundamental characteristic of recurrent neural
  networks},\ }\href@noop {} {\bibfield  {journal} {\bibinfo  {journal} {Neur.
  Comp.}\ }\textbf {\bibinfo {volume} {25}},\ \bibinfo {pages} {671} (\bibinfo
  {year} {2013})}\BibitemShut {NoStop}%
\bibitem [{\citenamefont {Takens}(2006)}]{takens2006detecting}%
  \BibitemOpen
  \bibfield  {author} {\bibinfo {author} {\bibfnamefont {F.}~\bibnamefont
  {Takens}},\ }\bibfield  {title} {\bibinfo {title} {Detecting strange
  attractors in turbulence},\ }in\ \href@noop {} {\emph {\bibinfo {booktitle}
  {Dynamical Systems and Turbulence, Warwick 1980: proceedings of a symposium
  held at the University of Warwick 1979/80}}}\ (\bibinfo {organization}
  {Springer},\ \bibinfo {year} {2006})\ pp.\ \bibinfo {pages}
  {366--381}\BibitemShut {NoStop}%
\bibitem [{\citenamefont {McCann}\ and\ \citenamefont
  {Yodzis}(1994)}]{McCann1994}%
  \BibitemOpen
  \bibfield  {author} {\bibinfo {author} {\bibfnamefont {K.}~\bibnamefont
  {McCann}}\ and\ \bibinfo {author} {\bibfnamefont {P.}~\bibnamefont
  {Yodzis}},\ }\bibfield  {title} {\bibinfo {title} {Nonlinear dynamics and
  population disappearances},\ }\href {https://doi.org/10.1086/285714}
  {\bibfield  {journal} {\bibinfo  {journal} {Am. Nat.}\ }\textbf {\bibinfo
  {volume} {144}},\ \bibinfo {pages} {873–879} (\bibinfo {year}
  {1994})}\BibitemShut {NoStop}%
\bibitem [{\citenamefont {Bergstra}\ \emph {et~al.}(2011)\citenamefont
  {Bergstra}, \citenamefont {Bardenet}, \citenamefont {Bengio},\ and\
  \citenamefont {K\'{e}gl}}]{NIPS2011}%
  \BibitemOpen
  \bibfield  {author} {\bibinfo {author} {\bibfnamefont {J.}~\bibnamefont
  {Bergstra}}, \bibinfo {author} {\bibfnamefont {R.}~\bibnamefont {Bardenet}},
  \bibinfo {author} {\bibfnamefont {Y.}~\bibnamefont {Bengio}},\ and\ \bibinfo
  {author} {\bibfnamefont {B.}~\bibnamefont {K\'{e}gl}},\ }\bibfield  {title}
  {\bibinfo {title} {Algorithms for hyper-parameter optimization},\ }in\ \href
  {https://proceedings.neurips.cc/paper_files/paper/2011/file/86e8f7ab32cfd12577bc2619bc635690-Paper.pdf}
  {\emph {\bibinfo {booktitle} {Advances in Neural Information Processing
  Systems}}},\ Vol.~\bibinfo {volume} {24},\ \bibinfo {editor} {edited by\
  \bibinfo {editor} {\bibfnamefont {J.}~\bibnamefont {Shawe-Taylor}}, \bibinfo
  {editor} {\bibfnamefont {R.}~\bibnamefont {Zemel}}, \bibinfo {editor}
  {\bibfnamefont {P.}~\bibnamefont {Bartlett}}, \bibinfo {editor}
  {\bibfnamefont {F.}~\bibnamefont {Pereira}},\ and\ \bibinfo {editor}
  {\bibfnamefont {K.}~\bibnamefont {Weinberger}}}\ (\bibinfo  {publisher}
  {Curran Associates, Inc.},\ \bibinfo {year} {2011})\BibitemShut {NoStop}%
\bibitem [{\citenamefont {Bergstra}\ and\ \citenamefont
  {Bengio}(2012)}]{Bergstra2012}%
  \BibitemOpen
  \bibfield  {author} {\bibinfo {author} {\bibfnamefont {J.}~\bibnamefont
  {Bergstra}}\ and\ \bibinfo {author} {\bibfnamefont {Y.}~\bibnamefont
  {Bengio}},\ }\bibfield  {title} {\bibinfo {title} {Random search for
  hyper-parameter optimization.},\ }\href@noop {} {\bibfield  {journal}
  {\bibinfo  {journal} {J. Mach. Learn. Res.}\ }\textbf {\bibinfo {volume}
  {13}} (\bibinfo {year} {2012})}\BibitemShut {NoStop}%
\bibitem [{\citenamefont {Zoph}\ and\ \citenamefont {Le}(2016)}]{Zoph2016}%
  \BibitemOpen
  \bibfield  {author} {\bibinfo {author} {\bibfnamefont {B.}~\bibnamefont
  {Zoph}}\ and\ \bibinfo {author} {\bibfnamefont {Q.~V.}\ \bibnamefont {Le}},\
  }\bibfield  {title} {\bibinfo {title} {Neural architecture search with
  reinforcement learning},\ }\href@noop {} {\bibfield  {journal} {\bibinfo
  {journal} {arXiv preprint arXiv:1611.01578}\ } (\bibinfo {year}
  {2016})}\BibitemShut {NoStop}%
\bibitem [{\citenamefont {Williams}\ and\ \citenamefont
  {Rasmussen}(2006)}]{williams2006}%
  \BibitemOpen
  \bibfield  {author} {\bibinfo {author} {\bibfnamefont {C.~K.}\ \bibnamefont
  {Williams}}\ and\ \bibinfo {author} {\bibfnamefont {C.~E.}\ \bibnamefont
  {Rasmussen}},\ }\href@noop {} {\emph {\bibinfo {title} {Gaussian Processes
  for Machine Learning}}},\ Vol.~\bibinfo {volume} {2}\ (\bibinfo  {publisher}
  {MIT press Cambridge, MA},\ \bibinfo {year} {2006})\BibitemShut {NoStop}%
\bibitem [{\citenamefont {Breiman}(2001)}]{Breiman2001}%
  \BibitemOpen
  \bibfield  {author} {\bibinfo {author} {\bibfnamefont {L.}~\bibnamefont
  {Breiman}},\ }\bibfield  {title} {\bibinfo {title} {Random forests},\
  }\href@noop {} {\bibfield  {journal} {\bibinfo  {journal} {Mach. Learn.}\
  }\textbf {\bibinfo {volume} {45}},\ \bibinfo {pages} {5} (\bibinfo {year}
  {2001})}\BibitemShut {NoStop}%
\end{thebibliography}

%
\end{document}